\documentclass{aa}
\usepackage{txfonts}
\usepackage[sort]{natbib}
\bibpunct{(}{)}{;}{a}{}{,}  
\usepackage{graphicx}
\usepackage{txfonts}

\def\Ms{{M$_{\sun}$}}
\def\MKo{{M$_\mathrm{K_0}$}}
\def\MK{{M$_\mathrm{K}$}}
\def\AK{{A$_\mathrm{K}$}}
\def\AV{{A$_\mathrm{V}$}}
\def\Ks{{K$_\mathrm{s}$}}
\def\Qs{{Q$_\mathrm{s}$}}
\def\Qd{{Q$_\mathrm{d}$}}
\def\th{{$^\mathrm{h}$}}
\def\tm{{$^\mathrm{m}$}}
\def\Brg{{Br$_\gamma$}}
\def\SFRa{{$\Sigma_\mathrm{SFR}$}}

\def\msyr{{M$_{\sun}$ yr$^{-1}$}}

\begin{document}
\title{Star formation in grand-design, spiral galaxies\thanks{Based
    on observations collected at the European Southern Observatory, Chile;
    program: ESO 82.B-0331}}
\subtitle{Young, massive clusters in the near-infrared}
\titlerunning{Star formation in spiral galaxies}
\author{P.~Grosb{\o}l\inst{\ref{inst1}} \and H.~Dottori\inst{\ref{inst2}}}
\offprints{P.~Grosb{\o}l}
\institute{European Southern Observatory,
  Karl-Schwarzschild-Str.~2, 85748 Garching, Germany \\
  \email{pgrosbol@eso.org}\label{inst1}
\and
  Instituto de F\'{i}sica, Univ. Federal do Rio Grande do Sul,
  Av. Bento Gon\c{c}alves~9500, 91501-970 Porto Alegre, RS, Brazil \\
  \email{dottori@ufrgs.br}\label{inst2}
}

\date{Received 15 September 2011 / Accepted 23 February 2012}

\abstract{} 
{Spiral structure is a prominent feature in many disk galaxies and is often
  outlined by bright, young objects.  We study the distribution of young
  stellar clusters in grand-design spiral galaxies and thereby determine
  whether strong spiral perturbations can influence star formation.  }
{Deep, near-infrared JHK-maps were observed for ten nearby, grand-design,
  spiral galaxies using HAWK-I at the Very Large Telescope.  Complete,
  magnitude-limited candidate lists of star-forming complexes were obtained by
  searching within the K-band maps.  The properties of the complexes were
  derived from (H-K)--(J-H) diagrams including the identification of the
  youngest complexes (i.e. $\la$7\,Myr) and the estimation of their
  extinction. }
{Young stellar clusters with ages $\la$7\,Myr have significant internal
  extinction in the range of \AV=3-7\tm, while older ones typically have
  \AV$<$ 1\tm.  The cluster luminosity function (CLF) is well-fitted by a
  power law with an exponent of around -2 and displays no evidence of a high
  luminosity cut-off.  The brightest cluster complexes in the disk reach
  luminosities of \MK\ = -15\fm5 or estimated masses of 10$^6$\Ms.  At radii
  with a strong, two-armed spiral pattern, the star formation rate in the arms
  is higher by a factor of 2-5 than in the inter-arm regions.  The CLF in the
  arms is also shifted towards brighter \MK\ by at least 0\fm4.  We also
  detect clusters with colors compatible with Large Magellanic Cloud
  intermediate age clusters and Milky Way globular clusters.  The 
  (J-K)-\MK\ diagram of several galaxies shows, for the brightest clusters,
  a clear separation between young clusters that are highly attenuated by
  dust and older ones with low extinction. }
{The gap in the (J-K)-\MK\ diagrams implies that there has been a rapid
  expulsion of dust at an age around 7\,Myr, possibly triggered by supernovae.
  Strong spiral perturbations concentrate the formation of clusters in the arm
  regions and shifts their CLF towards brighter magnitudes.  }

\keywords{galaxies:~spiral -- galaxies:~structure --
  galaxies:~star~clusters:~general --  galaxies: star formation --
  infrared:~galaxies -- techniques:~photometry} 
\maketitle

\section{Introduction}
A large fraction of the star formation in the local universe occurs in
gas-rich, disk galaxies as indicated by the Schmidt-Kennicutt law
\citep{schmidt59, kennicutt89}, which relates the star formation rate (SFR) to
the surface/volume density of gas at kpc scales.  The influence of spiral
structure on the global SFR of galaxies is limited as indicated by the small
difference seen between multi-armed and grand-design spirals
\citep{elmegreen86}.  On smaller scales, the relation is less clear, partly
owing to the low spatial resolution of gas maps \citep{bigiel08}, and may be
influenced by local features such as resonances, shear, and spiral structures.
This may indicate that the main effect of spiral structure is to concentrate
the SFR into the arms rather than to change the global rate
\citep{kennicutt98}.

\citet{seigar02b} analyzed the SFR of 20 spiral galaxies, concluding that the
SFR is enhanced significantly in the vicinity of their K-band arms and
correlates well with the relative strength of shocks in arms.  This led them
to suggest that density waves trigger star formation in the vicinity of spiral
arms.

The star formation in the arm and inter-arm regions of three spirals was
studied by \citet{foyle10} using data from the ultraviolet to radio.  They
found that star formation in the inter-arm regions was significant and only
marginally enhanced in the arms.

\begin{table*}
 \caption[]{List of galaxies. Name and Hubble type are listed as given in RSA.
   Filter bands observed, seeing in the final, stacked K-band image expressed
   as the full-width at half-maximum ({\tt fwhm}) in arcsec, and \Ks\ surface
   brightness $\mu_\mathrm{5}$ in magnitudes per pixel for which a
   signal-to-noise ratio of five was reached are also shown. The adopted
   distance modulus, m-M, distance D in Mpc, and the corresponding linear
   resolution $\Delta$s in pc are provided.  The position angle PA and
   inclination angle IA adopted for the galaxies are listed in degrees.
   Limiting magnitudes K$_l$, H$_l$, J$_l$, and Y$_l$ are indicated
   corresponding to a completeness of around 90\% (see text).  Finally, the
   total number of sources N$_s$ in the galaxy disks for which aperture
   magnitudes could be estimated is given.}
 \label{tbl:gal}
 \begin{tabular}{lllcrrrrrrrrrrr}
\hline\hline
  Galaxy & Type & Bands &{\tt fwhm} & \multicolumn{1}{c}{$\mu_\mathrm{5}$} &
  \multicolumn{1}{c}{m-M} & \multicolumn{1}{c}{D} &
  \multicolumn{1}{c}{$\Delta$s} &
  \multicolumn{1}{c}{PA} & \multicolumn{1}{c}{IA} &
  \multicolumn{1}{c}{K$_l$} & \multicolumn{1}{c}{H$_l$} &
  \multicolumn{1}{c}{J$_l$} & \multicolumn{1}{c}{Y$_l$} &
  \multicolumn{1}{c}{N$_s$}  \\ \hline
    \object{NGC~157}  &  Sc(s) I-II  &  YJHK      & 0.37 & 23.9 &
      31.28 & 18.0 & 32 &
      40\tablefootmark{a} & 45\tablefootmark{a} &
      20.19 & 20.71 & 21.27 & 21.75 & 2254 \\
    \object{NGC~1232} &  Sc(rs) I    &  YJHK      & 0.38 & 24.1 &
      31.48 & 19.8 & 36 &
      90\tablefootmark{b} & 30\tablefootmark{b} & 
      20.62 & 20.80 & 21.72 & 22.21 & 3177 \\
    \object{NGC~1300} &  SBb(s) I.2  &  YJHK      & 0.55 & 24.0 &
      31.46 & 19.6 & 52 &
      87\tablefootmark{c} & 35\tablefootmark{c} & 
      19.80 & 20.30 & 21.18 & 21.30 &  614 \\
    \object{NGC~1365} &  SBb(s) I    &  YJHK      & 0.40 & 23.7 &
      31.62 & 21.1 & 41 &
      40\tablefootmark{d} & 40\tablefootmark{d} & 
      20.17 & 20.32 & 21.24 & 21.73 & 2417 \\
    \object{NGC~1566} &  Sc(s) I     &  YJHK      & 0.44 & 23.1 &
      31.55 & 20.5 & 44 &
      41\tablefootmark{e} & 27\tablefootmark{e} & 
      19.29 & 19.82 & 20.64 & 21.18 &  969 \\
    \object{NGC~2997} &  Sc(s) I.3   &  YJHK\Brg  & 0.38 & 23.8 &
     31.42 & 19.2 & 35 &
     107\tablefootmark{f} & 32\tablefootmark{f} & 
     20.14 & 20.33 & 21.27 & 21.75 & 5313 \\
    \object{NGC~4030} &  Sbc(r) I    &  YJHK      & 0.44 & 23.6 &
      31.99 & 25.0 & 53 &
      27\tablefootmark{g} & 44\tablefootmark{g} & 
      19.70 & 20.19 & 20.76 & 21.26 & 1196 \\
    \object{NGC~4321} &  Sc(s) I     &  YJHK      & 0.66 & 23.7 &
     32.07 & 26.0 & 83 &
     153\tablefootmark{h} & 27\tablefootmark{h} & 
     19.32 & 19.81 & 20.74 & 21.66 & 1184 \\
    \object{NGC~5247} &  Sc(s) I-II  &  YJHK      & 0.41 & 23.7 &
      31.77 & 22.6 & 45 &
      97\tablefootmark{h} & 29\tablefootmark{h} & 
      19.79 & 20.65 & 21.26 & 21.75 & 2259 \\
    \object{NGC~7424} &  Sc(s) II.3  &  JHK       & 0.41 & 24.3 &
      29.88 &  9.5 & 19 &
      44\tablefootmark{i} & 27\tablefootmark{i} & 
      20.80 & 21.23 & 22.21 &   -   & 6137 \\
  \hline   
\end{tabular}
\tablefoot{References to the projection parameters adopted:
  \tablefoottext{a}{\citet{zurita02}},
  \tablefoottext{b}{\citet{vanzee99}},
  \tablefoottext{c}{\citet{lindblad97}},
  \tablefoottext{d}{\citet{pence90}},
  \tablefoottext{e}{\citet{hess09}},
  \tablefoottext{f}{\citet{ganda06}},
  \tablefoottext{g}{\citet{knapen93}},
  \tablefoottext{h}{\citet{kuno07}}, and
  \tablefoottext{i}{\citet{becker88}}.
}
\end{table*}

A study of cluster populations in 21 nearby spiral galaxies led
\citet{larsen99} to conclude that the age distribution of clusters in disk
galaxies shows no obvious peak, indicating that massive clusters are formed as
part of an ongoing process rather than in bursts.  They also concluded that
the radial distribution of clusters follows that of the H$_\alpha$ surface
brightness.  Hubble Space Telescope (HST) high resolution observations
\citep{larsen09} suggested that the initial cluster mass function in
present-day spiral disks can be modeled as a Schechter function
\citep{schechter76} with a cut-off mass M$_c\approx\,2x10^5$\Ms.

There are only a small number of grand-design spirals with prominent arms at a
distance where individual clusters can be analyzed.  One such galaxy,
NGC~5194, was studied by \citet{scheepmaker07, scheepmaker09}. These authors
found that the spatial distribution of the star clusters younger than 10\,Myr
shows the strongest correlation with the spiral arms, H$_\alpha$, and radio
continuum emission, and that this correlation decreases with age.

Most of these studies are based on H$_\alpha$, UV, and visual broad-band
observations, which are significantly biased by extinction.  In spiral arms
with prominent dust lanes, it is difficult to conduct a complete census of
star forming regions and cluster complexes in the visible especially for the
youngest ones, which may still be embedded in dust.  In these cases, an almost
unbiased, magnitude-limited sample can be obtained more readily in the
near-infrared (NIR), where the attenuation by dust is much smaller (e.g.
\AK$\approx$ 0.1~\AV).

In general terms, the K-band light from the disks of spiral galaxies is
dominated by an old stellar population \citep{rix93a}. Nevertheless, in a
study of the NIR K-band images of spiral galaxies \citet{gp98} noted that
several grand-design spirals have bright knots along their arms.  The
alignment and concentration of these knots in the arms suggested that they
were associated with young objects.  This was confirmed by the NIR spectra of
several knots in NGC~2997 \citep{grosbol06} displaying strong \Brg\ emission,
which demonstrates that the knots are \ion{H}{II} regions deeply enshrouded in
dust.  Both \Brg\ emission and NIR colors can be used as age indicators for
sources younger than $\sim$10\,Myr \citep[ hereafter SB99]{leitherer99} and
allow the study of the early phases of star formation, which are often hidden
by dust.

A sample of 46 spiral galaxies was studied by \citet{grosbol08}, who found
that around 70\% of these grand-design spirals have a significant
concentration of bright K-band knots in their arms.  The NIR color-color
diagrams of four of the galaxies with JHK photometry suggest that a
significant fraction of the extended sources are complexes of young stellar
clusters with ages $\lesssim$10\,Myr that are reddened by several magnitudes
of visual extinction. The brightest knots reach \MK\ = -15\fm5 corresponding
to stellar clusters or complexes with total masses of up to several $10^5$\Ms,
accounting for the typical SFR of 1\,\Ms\ yr$^{-1}$ measured for the brightest
grand-design spiral galaxies.

The main emphasis of our paper is to characterize the general distribution of
clusters in grand-design spiral galaxies.  Relying on NIR photometry, this
study intends to provide a more complete sampling than in visual bands
including highly reddened, young clusters in the arm regions.  Our access to
younger, partly embedded clusters allows us to estimate of extinction at early
phases of star formation and thereby establish a clearer definition of the
CLF.  It is still unclear whether spiral structure can influence star
formation on small scales \citep[see e.g.][]{bigiel08, seigar02b}.  The
combination of NIR colors and spatial information offers a detailed view of
the distribution of clusters as a function of their location relative to
spiral arms.  This is used to test whether strong spiral perturbations can
affect local star formation.

A long-term objective, to be pursued in a forthcoming paper, is the study of
the possible implications of the spatial distribution of newly formed clusters
on the nature of spiral arms.  The ability to estimate the individual ages of
the youngest clusters \citep{grosbol09} could reveal spatial age gradients
that can be compared with predictions for different spiral arms scenarios.  It
should be possible to distinguish between density wave \citep{lin64, bertin89,
  roberts69a, gittins04} and material arms \citep{wada11} owing to their
different relative angular velocities.  In addition, the contrast between star
formation in arm and inter-arm regions \citep{seigar98, seigar02b} could be
used as an indicator of the ongoing mechanism. However, owing to the intrinsic
velocity dispersion of clusters and the limited age resolution, the current
data cannot be used to estimate the longevity of spirals, that is whether they
are long-lived quasi-stationary spiral structures resulting from density waves
\citep{fujii11} or transient spiral patterns that occur because of recurrent
gravitational instabilities \citep{foyle11, sellwood10, sellwood11}.

In the current paper, we first present the NIR data and describe the
identification of the stellar cluster complexes and extraction of their
properties.  Color-color and color-magnitude diagrams of the cluster complexes
are presented in Sect.~\ref{sec:ccd} with a discussion of the implications for
their physical properties.  Star formation rates and CLFs are derived in
Sects.~\ref{sec:clf} and \ref{sec:sfr}.  The general distribution of cluster
complexes in the arm and inter-arm regions is given in
Sect.~\ref{sec:spacial}.  The appendices includes descriptions, maps, and
diagrams for the individual galaxies.

\begin{figure*}
  \resizebox{\hsize}{!}{\includegraphics{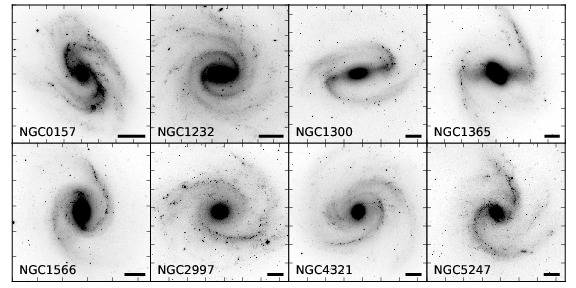}}
  \caption{Direct \Ks-maps of the eight most prominent grand-design spiral
    galaxies in the sample.  The images are presented in a negative, arbitrary
    intensity scale to more clearly display their spiral structure.  All
    images are orientated with north to the top and east to the left with a
    scale indicated by the 30\arcsec\ bar in the lower right corner. }
  \label{fig:gal}
\end{figure*}

\section{Data and reduction}
\label{sec:data}
Eight grand-design galaxies with relatively high SFRs and concentrations of
stellar clusters in their arms were selected from the list of
\citet{grosbol08}.  The candidates were selected to have inclination angles IA
$<$ 45\degr\ and a geometry that made it possible to map their spiral
structure unambiguously.  To get a reasonable range of morphological types
with strong spiral perturbations, including bars, it was necessary to include
galaxies with distances up to around 25~Mpc.  Two grand-design spirals with
weak spiral arms (i.e. NGC~1232 and NGC~7424) were added to allow a comparison
of properties as a function of perturbation strength.  The galaxy NGC~7424 has
an estimated distance of 10~Mpc, which permits a study of its faint clusters
and therefore a more reliable estimate of the shape of the cluster
distribution at low masses.  The galaxies are listed in Table~\ref{tbl:gal}
where their morphological types according to \citet[ hereafter RSA]{sandage81}
are also given.  Although only two galaxies are classified as strongly barred,
all the other galaxies display weak bars or oval distortions in their central
regions on their K-band maps. The eight main grand-design galaxies in the
sample are shown in Fig.~\ref{fig:gal}.  All the galaxies are displayed in
Appendix~\ref{app:gal}, where a short description of their morphology is also
given.

The galaxies were observed with the HAWK-I/VLT instrument, which employs a
mosaic of 4 Hawaii 2RG 2048$\times$2048 chips with a total field of
7\arcmin\ and a pixel size of 0.106\arcsec.  All observations were performed
in service mode except for NGC~7424, which was acquired during the HAWK-I
commissioning.  We designed a single observing block for each galaxy,
including the 4 broad-band filters, in service mode to ensure that the
observing conditions were as similar as possible.  Target and sky exposures
were interleaved with a maximum duration of 3\tm\ between empty field
exposures to allow accurate sky subtraction.  Typical total exposure times on
source were for \Ks\ in the range 9-12\tm, while they were 7\tm, 4\tm, and
4\tm\ for the filters H, J, and Y, respectively. In addition, a \Brg-frame
with a total exposure of 24\tm\ on source was secured for NGC~2997.  The
individual frames were offset to ensure that all parts of the galaxies were
exposed including the 15\arcsec\ gaps between the chips. A minimum of four
frames were taken for the J- and Y-bands which required shorter exposure
times.  In these cases, parts of the field were covered by only one or two
exposures, reducing the ability to reject bad pixels and cosmic ray events
there.

A dedicated, ESO-MIDAS based pipeline was used for the data reduction because
a generic template, which is not supported by the standard ESO pipeline, was
used for the observation.  The observatory-provided dark and flat-field frames
were used.  Since the flats were taken during the morning and evening, they
may contain small gradients.  A low-order, polynomial correction was applied
to the observatory flats estimated from stacked sky exposures obtained during
the nights.  The sky frames were inspected visually and rejected if odd
features were seen.  Furthermore, all discrete sources detected on them were
removed.  The average of adjacent sky exposures was then subtracted from the
target frames.  The resulting frames had to be re-sampled before stacking
because the different detector chips were not perfectly aligned.  A nearest
pixel algorithm was applied since the pixel size was significantly smaller
than the average seeing of the frames and an interpolation would smear out bad
pixels.  The transformation constants including gains for the individual chips
were taken from the instrument manual and applied.

\begin{figure}
  \resizebox{\hsize}{!}{\includegraphics{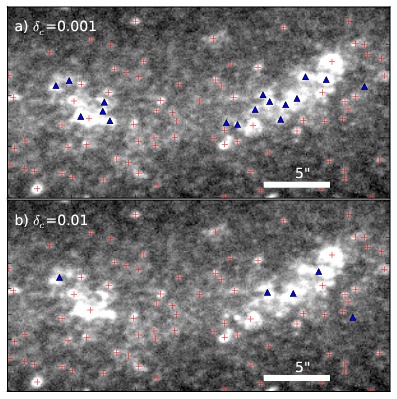}}
  \caption{Sources detected in a star forming complex in the southern arm of
    NGC~2997 ($\alpha_{2000}$= 9\th 45\tm 33, $\delta_{2000}$ = -31\degr
    21\arcmin 28\arcsec) using different de-blending contrast thresholds (a)
    $\delta_c=0.001$ and (b) $\delta_c=0.01$. Sources with identical positions
    for both $\delta_c$ values are indicated by `+', while differences are
    marked by `$\bigtriangleup$'. }
  \label{fig:sex}
\end{figure}

The galaxy frames were aligned to within one pixel using stars in the field
and centered on the K-band intensity peak of their bulges.  The quality of the
individual frames was checked by comparing the magnitude and seeing of a
bright field star, which in a few cases led to the rejection of frames.  Bad
pixels and outliers could be identified when at least three frames were
available.  These pixels were omitted when their inclusion would have
significantly increased the variance around the mean \citep{grosbol04}. Only
for NGC~1566, observed in rather variable conditions, was it necessary to
remove a frame from the J and Y stacks.  This left an unexposed square of
15$\times$15\arcsec\ close to the center of NGC~1566 in the J- and Y-maps.

The HAWK-I field size allowed the use of foreground stars from 2MASS \citep[
  hereafter 2MASS]{2mass} for the photometric calibration.  The uncertainty in
the photometric zero points was smaller than 0\fm08 for all maps based on 7-60
2MASS stars depending on Galactic latitude of the fields.  For the Y-band,
2MASS magnitudes were estimated from the transformation by \citet{hodgkin09}
for the WFCAM instrument with similar filters.  Using these 2MASS stars,
astrometric transformations were also derived with typical errors of
$<$0.2\arcsec.  The full-width at half-maximum ({\tt fwhm}) of stellar images
on the final, stacked \Ks-maps is listed in Table~\ref{tbl:gal} where the
surface brightness $\mu_\mathrm{5}$ for a signal-to-noise ratio of five
per pixel is also given.  The other bands had similar seeing and surface
brightness limits.  All non-stellar sources were corrected for foreground
absorption due to our Galaxy using the maps of \citet{schlegel98} provided by
the NASA/IPAC Extragalactic Database (NED).

The distance moduli for the galaxies were derived from their systemic
velocities relative to the 3K CMB using a Hubble constant of 73 km s$^{-1}$
Mpc$^{-1}$ as given by NED.  They differ significantly in several cases from
those obtained using the Galactic standard-of-rest \citep[ hereafter
  RC3]{rc3}, such as for NGC~2997 for which the distance adopted places the
galaxy almost 60\% further away.  The 3K~CMB corrections were preferred since
they gave more consistent magnitudes for the brightest sources in the
galaxies.  The distances adopted are listed in Table~\ref{tbl:gal} together
with the linear resolution corresponding to the {\tt fwhm}.  For most of the
galaxies, sources with a size of 50~pc could be resolved.  This does not allow
us to identify individual stellar clusters but rather complexes of clusters and
star forming regions.  The scale is compatible with that of typical giant
molecular cloud (GMC) sizes suggesting that the sources detected are
associated with individual GMCs.

\begin{figure}
  \resizebox{\hsize}{!}{\includegraphics{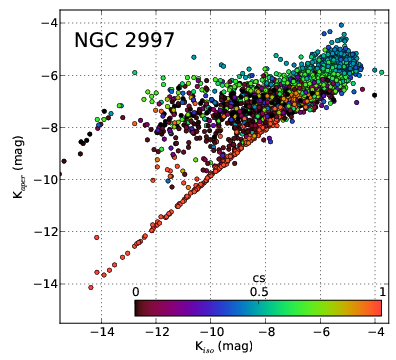}}
  \caption{Instrumental aperture magnitudes for sources detected in the
    \Ks-band image of NGC~2997 as a function of their isophotal magnitudes.
    Only stars with photometric errors $<$0\fm2 are plotted. The color
    indicates the value of $cs$ from 1 for stars (red) to 0 for diffuse
    objects (blue).}
  \label{fig:iso}
\end{figure}

\begin{figure*}
  \resizebox{\hsize}{!}{\includegraphics{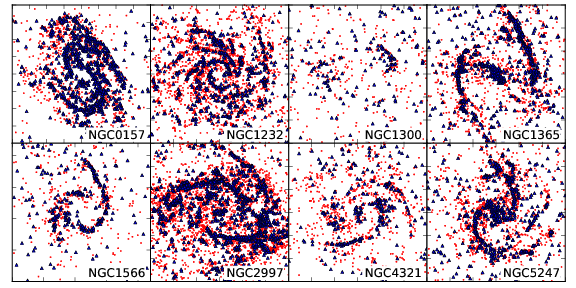}}
  \caption{Locations of non-stellar sources with errors $<$0\fm3 identified in
    the galaxies shown in Fig.~\ref{fig:gal}.  Compact sources ($0.3<cs<0.95$)
    are indicated by dots (red) while diffuse ones ($cs<0.3$) are plotted as
    triangles (blue).}
  \label{fig:cxy}
\end{figure*}

\section{Detection of sources}
\label{sec:sex}
We performed our search for sources with {\it SExtractor} v.2.8.6
\citep{bertin96} on the K-band frames since they are less affected by dust
extinction and are therefore expected to yield a less biased list of
candidates.  The steep flux gradients in the central parts of the galaxies
significantly limited the detection of faint sources in these regions.  To
remove these gradients, a smoothed version of the K images was subtracted
before applying the search.  The smoothed image can be generated by using
either a parametric model fitted to the surface brightness distribution of the
galaxy or digital filtering.  The latter option was preferred to avoid any
bias being caused by the model adopted.  The filter consisted first of a
median filter with a kernel of 5\arcsec\ squared followed by a Gaussian
smoothing to reduce the noise level.  This retained all sources with
significant gradients on scales smaller than the filter size, while removing
the smooth variation due to the spiral arms, exponential disk, and bulge,
except in the central 5-10\arcsec\ of the galaxies.  It allowed the use of a
large background mesh for {\it SExtractor}.  The search was performed with a
convolution mask of a {\tt fwhm} of two pixels corresponding to half of the
typical seeing.  A detection threshold of 2.5 times the background noise was
used and only sources with at least four pixels were accepted.

Several large and complex sources exist along the spiral arms.  They were
occasionally separated into individual objects depending on the {\it
  SExtractor} de-blending contrast threshold parameter $\delta_c$.  A typical
complex region in the southern arm of NGC~2997 is displayed in
Fig.~\ref{fig:sex}, where sources detected with two different values of
$\delta_c$ are shown.  Without the resolution to resolve individual stellar
clusters, the choice of the 'best' value for $\delta_c$ is subjective.
Comparing several different complex regions, $\delta_c=0.001$ was found to
yield a reasonable separation and was used for the final search.

The {\it class\_star} estimator, $cs$, provided by {\it SExtractor} was used
to indicate the shape of sources where $cs=1$ corresponds to star-like images
and $cs=0$ to diffuse ones.  Sources with $cs>0.95$ were assumed to be
foreground stars. The distributions of non-stellar sources in the eight main
targets are shown in Fig~\ref{fig:cxy}.  It is clear that the diffuse sources
($cs<0.3$) outline the spiral structure more clearly than the compact ones,
which are more randomly distributed within the galaxies.  Only for two of the
galaxies is this not the case, namely for NGC~1300 where the main star
formation occurs at the end of its bar, and for NGC~7424, which has a weak
spiral perturbation.  All the galaxies are displayed in Fig.~\ref{figa:cxy}.

The centroids of the candidates found by this procedure were then used to
calculate the magnitudes in each available band.  Both aperture and isophotal
magnitudes were estimated since they have somewhat different properties and do
not assume a specific shape of the source.  The former was based on an
aperture with a diameter of 1\arcsec\ and a background area consisting of a
2\arcsec\ ring with a radius of 4\arcsec, to which a $\kappa\sigma$-clipping
filter was applied.  Isophotal magnitudes were computed by mapping the
\Ks-band pixels across a region of 4\arcsec\ around the centroid into either
`source' or `background' pixels.  This was done by moving out from the local
maximum near the centroid in a spiral pattern, while monitoring the pixel
values.  The background level was estimated as the minimum value within the
region for which no significant gradient could be determined.  Using this
level and the associated standard deviation, the pixels were classified as
`source' if they were within the {\tt fwhm} of or $3\sigma$ above the
background, while all pixels consistent with the level were flagged as
'background'.  The isophotal magnitude of a source was computed using this map
for all the bands, which ensured that the same set of pixels was used to
calculate color indices.  We insisted that a source had more `background' than
'source' pixels.

The two types of magnitudes are compared in Fig.~\ref{fig:iso}, where they are
shown for objects detected in NGC~2997.  Bright, stellar-like sources show a
good correlation with a slope of unity.  By definition, isophotal magnitudes
do not record the same fraction of flux at all magnitude levels.  This effect
may approach 0\fm5 for the faintest magnitude used in this sample.  Aperture
magnitudes may also be brighter when other sources, in crowded fields, are
included in the aperture.  The opposite trend is seen for brighter sources,
where many non-stellar objects are more luminous according to their isophotal
magnitude than that computed for a fixed aperture.  The effect appears for
instrumental magnitudes $<$-8\tm\ for which typical non-stellar objects have
more 'source' pixels than the aperture used (i.e. approx. 280 pixels).  This
suggests that the aperture estimate misses flux associated with the source
because of its fixed size. In general, aperture magnitudes give a more
reliable and consistent measure of the fluxes associated with the individual
sources and were used in the current study.  The photometric data for the
sources detected in the galaxies are available in digital form from
CDS\footnote{Centre de Donn{\'e}es astronomiques de Strasbourg:
  http://cds.u-strasbg.fr}.  These tables contain the aperture photometry for
each filter with associated estimates of the error and background.  In
addition, we provide the coordinates and $cs$ index. The estimation of the
total flux for the largest, most complex sources can be obtained more
accurately from isophotal magnitudes but remains uncertain owing to the
arbitrary division into substructures (i.e. choice of de-blending threshold
$\delta_c$).

For a typical galaxy in the sample with a disk of 10 arcmin$^2$, the
contribution of background galaxies is 97$\pm$12 sources with K$<$20\tm\ or
44$\pm$8 with K$<$19\tm\ using the estimates derived by the {\it GalaxyCount}
program \citep{ellis06}.  To limit the number of these galaxies in the sample,
only sources within the galactic disks were assumed to be defined
when their \Ks-band background surface brightness was 3$\sigma$ above sky.

The limiting magnitudes were estimated from the histogram of apparent
magnitudes as the number weighted average of the highest bin and its two
neighbors.  Comparing with simulated fields of sources with a power-law
luminosity function, the estimate corresponds to roughly 90\% completeness.
The total number of non-stellar sources for which aperture photometry could be
estimated is given in Table~\ref{tbl:gal}, where limiting magnitudes for
all filters are also listed.

\begin{figure}
  \resizebox{\hsize}{!}{\includegraphics{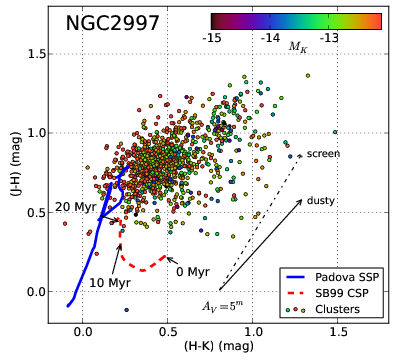}}
  \caption{Color-color diagram for non-stellar sources in NGC~2997 with errors
    $<$0\fm05.  Evolutionary tracks for a stellar population are shown where
    the full drawn line is for a Padova SSP model while the dashed line
    indicates a SB99 CSP model. The color coding of the points indicates their
    absolute magnitude \MK. A standard 'screen' reddening vector and one for a
    'dusty' cluster medium are shown for A$_\mathrm{V}$ = 5\tm. }
  \label{fig:cc2997}
\end{figure}

\section{Distribution of magnitudes and colors}
\label{sec:ccd}
A main tool for the study of the physical properties of the unresolved cluster
complexes is their integrated colors and magnitudes.  A typical (H-K)--(J-H)
color-color diagram (CCD) is shown in Fig.~\ref{fig:cc2997} for non-stellar
sources in NGC~2997 with photometric errors $<$0\fm05.  For reference, two
evolutionary tracks are shown for a stellar population with metallicity
Z=0.02, a Salpeter initial mass function (IMF) \citep{salpeter55}, and an
upper mass limit M$_u$ = 100\,\Ms.  The full drawn curve is a single-burst
stellar population (SSP) based on the Padova isochrones \citep{marigo08}
including the post thermal-pulse asymptotic-giant-branch (TP-AGB) phase. The
other track represents a continuous star formation population (CSP) from SB99
indicated by a dashed line.  The shift between the two tracks is mainly due to
the dust and nebular emission that is included in the SB99 models.  Even
though NIR extinction is much smaller than in the visual bands, it is still
significant.  As a first order compensation for attenuation by dust, one can
construct a reddening `corrected' color index Q assuming a linear correction
term. We consider two extreme cases: a screen model of the dust, which would
yield \Qs\ = (H-K) - 0.563$\times$(J-H) \citep{indebetouw05}, and a mix of
stars and dust giving \Qd\ = (H-K) - 0.844$\times$(J-H) \citep{witt92,
  israel98}. Both reddening vectors are indicated in Fig.~\ref{fig:cc2997} for
a visual absorption A$_\mathrm{V}$ = 5\tm.

Two main concentrations of sources can be identified in the CCD, namely a) one
of the complexes around ((H-K),(J-H)) = (0\fm4, 0\fm8), which corresponds to a
stellar population older than 20\,Myr and relatively low extinction, and b) a
weaker one close to (0\fm8, 1\fm1), which is compatible with younger and more
obscured complexes.  The remaining clusters are more homogeneously scattered
to higher (H-K). Their colors can be mainly modeled by SB99 adding specific
amounts of extinction and they are likely younger than 7\,Myr. However, to
model those clusters with the reddest (H-K), one would need to add a
significant amount of emission from hot dust.  We note that none of these
young complexes has a visual extinction of less than 2\tm.  Somewhat depending
on the reddening assumed, there is a tendency for younger sources to have a
higher extinction.

\begin{figure*}
  \resizebox{\hsize}{!}{\includegraphics{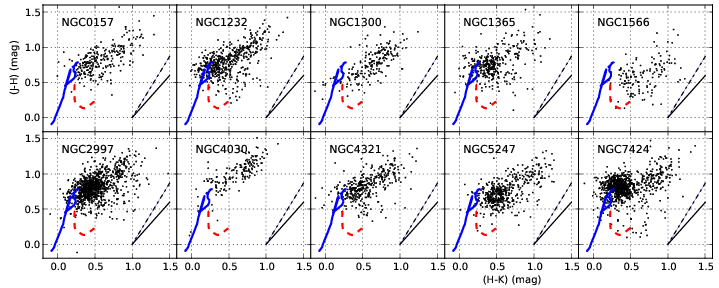}}
  \caption{Color-color diagrams for the galaxies.  Cluster evolutionary tracks
    and reddening vectors are shown for reference as in
    Fig.~\ref{fig:cc2997}.}
  \label{fig:ccall}
\end{figure*}

The color-color diagrams for all the galaxies are given in
Fig.~\ref{fig:ccall} for comparison.  Taking NGC~2997 as a reference, all the
CCDs share the same basic features, although with different emphasis.  The
position of the clump at the older end of the cluster tracks varies only
slightly among the galaxies with the exception of NGC~1566. In this galaxy, we
recognize the same morphology in the color-color diagram, but all complexes
are shifted almost 0\fm3 towards higher Q values.  This may partly result from
the variable weather conditions under which it was observed as mentioned in
Sect.~\ref{sec:data}.  The clump of older complexes contains a larger fraction
of faint sources, which makes its population size sensitive to the limiting
magnitude.  This effect can be observed by comparing the CCDs of NGC~2997
and NGC~4321, where the former has a 1\fm5 deeper limiting magnitude in
\MK. There are also intrinsic differences between the galaxies that can be
seen for galaxies with similar limiting absolute magnitudes, e.g. NGC~2997 has
more old complexes than NGC~1232, and NGC~4321 has more than NGC~4030.

The concentration of objects with higher (J-K) values, near (0\fm8, 1\fm2), is
present for all galaxies, although the gap to the older clump is most clearly
seen in NGC~2997, NGC~5247, and NGC~7424.  Comparing the relative position of
the two concentrations with the reddening vectors, it is evident that the
objects with higher (H-K) cannot be explained by older sources being
attenuated by dust since they would get higher (J-H) values than observed.
This shift can best be explained by the sources being younger in addition to
having high extinction.  The young, dusty objects in NGC~7424 are shifted even
more strongly to higher (H-K) values.  This could, in part, be due to the
higher linear resolution, which reduces the probability of several clusters
being included in one target aperture. Without spectroscopic data, it is
impossible to accurately estimate the physical properties of the complexes.
Variations in nebular emission and cluster IMF can shift the location of
sources in the CCD by at least 0\fm1.  Systematic differences in the amount of
bright, red, post TP-AGB stars (i.e. 2\tm\ $<$ (J-K)) between the clusters
could also play a role.  Full-scale CCDs for the galaxies are also shown in
Fig.~\ref{figa:ccall}.

\begin{figure}
  \resizebox{\hsize}{!}{\includegraphics{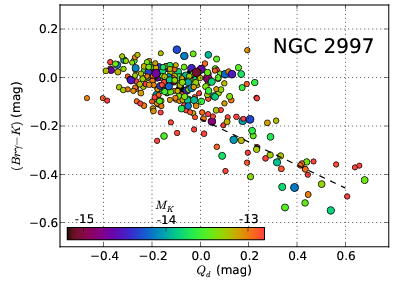}}
  \caption{\Brg\ index (\Brg-K) as a function of the \Qd\ index for
    non-stellar sources in NGC ~2997.  Sources with photometric errors in
    \Brg$<$0\fm03 are displayed with colors indicating \MK.  The dashed line
    represents the regression line. }
  \label{fig:brg}
\end{figure}

The evolutionary tracks from SB99 indicate that the youngest clusters display
strong \Brg\ and nebular emission.  The emission decreases with time and
vanishes around 7\,Myr in a way that depends slightly on the model parameters.
This suggests that the Q index, due to the emission in the \Ks-band, can be
used as an age indicator for clusters younger than 7\,Myr.  This was tested
using \Brg\ exposures of NGC~2997, which has a systemic velocity that permits
the HAWK-I \Brg-filter to be used.  The index (\Brg-K), with a zero point
defined by the foreground stars, is shown in Fig.~\ref{fig:brg} as a function
of the \Qd\ index.  The relation (\Brg-K) = -0.48 $\times$ \Qd\ - 0\fm17 is
determined, with a scatter of 0\fm08 for the 55 clusters with significant
\Brg-emission (i.e. (\Brg-K) $<$ -0\fm1) and errors in \Brg $<$0\fm03.  This
confirms that \Qd\ can be used to estimate the \Brg-emission and the age of
clusters with 0\fm1 $<$ \Qd\ for which the contamination of older clusters is
small. For \Qd$<$0\fm1, the index shows little variation and cannot be used as
an age indicator.  The equivalent width of \Brg\ in emission was previously
calibrated as an age indicator using SB99 models and K-band spectra
\citep{grosbol06}.

\begin{figure}
  \resizebox{\hsize}{!}{\includegraphics{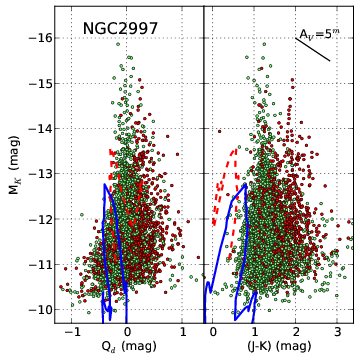}}
  \caption{Absolute K-band magnitudes for non-stellar sources in NGC~2997 with
    errors $<$0\fm3 as functions of \Qd\ and (J-K) indices. Padova and SB99
    SSP models for a cluster with 10$^5$\Ms\ are given by blue full-drawn and
    red dashed lines, respectively. The color coding indicates the other
    index, where red corresponds to 1\fm5$<$(J-K) and 0\fm1$<$\Qd,
    respectively.  The `dusty' reddening line is shown for A$_V$=5\tm.}
  \label{fig:cm2997}
\end{figure}

Color-magnitude diagrams (CMD) for NGC~2997 are given in Fig.~\ref{fig:cm2997}
using the \Qd\ and (J-K) color indices, which are roughly orthogonal.  For
clusters with \Qd$>$0\fm1, \Qd\ indicates age, whereas (J-K) relates to
extinction.  An evolutionary track for Padova and SB99 SSP models with
10$^5$~\Ms\ mass is shown for reference.  The exact location of the track
depends on both the upper mass limit M$_u$ and the upper slope of the IMF,
while changes in the metallicity around the solar value are of little
importance.  The upper limit to \Qd\ is closer to 0\fm5, which corresponds
well to the SB99 models with M$_u>$ 60~\Ms. The relatively few complexes with
higher \Qd\ values are likely affected by emission from hot dust.  There is a
tail of faint sources reaching negative values of \Qd\ around -0\fm5.  Many of
these are probably globular clusters that are similar to Galactic ones, which
have an average \Qd~= -0\fm4.  The \Qd\ distribution is narrower for brighter
complexes partly due to their higher average extinction implied by the CCD.
The brightest complexes have \MK\ reaching around -15\fm5, which corresponds
to masses close to 10$^6$~\Ms\ assuming a standard SB99 SSP model.

The right panel of Fig.~\ref{fig:cm2997} displays the (J-K) index, which shows
more clearly than (H-K) the effects of extinction.  For faint sources, the
lower boundary of (J-K) is sharp, in good agreement with the tracks for old
clusters with \AV$<$1\tm. The youngest sources dominate the high (J-K) values
and display a significant scatter due to variable extinction.  They form a
well-defined branch with (J-K) $\approx$ 1\fm8 that is clearly separated from
the bright, older sources with (J-K) around 1\fm2.  The model tracks predict
that youngest clusters should have (J-K) close to 0\fm0 and then approach
values in the range of 0.5-1\tm\ when massive blue stars evolve to red
super-giants.  If the extinction varied only slowly with age, one would not
expect there to be a separation between the young and old complexes in two
branches. Thus, the gap suggests that there has been a rapid reduction of
extinction, which could happen when the first supernovae explode and dust is
expelled from the cluster environment \citep{lada03, bastian06, goodwin06}.

\begin{figure}
  \resizebox{\hsize}{!}{\includegraphics{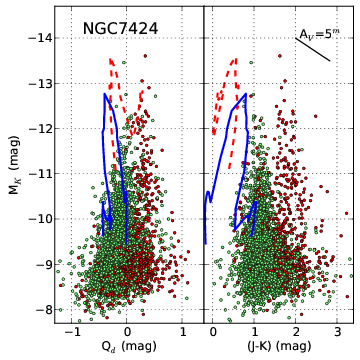}}
  \caption{Absolute K-band magnitudes for non-stellar sources in NGC~7424 with
    errors $<$0\fm3 as a function of \Qd\ and (J-K) indices. Colors and lines
    are as in Fig~\ref{fig:cm2997}. }
  \label{fig:cm7424}
\end{figure}

The CMDs for NGC~7424 are presented in Fig.~\ref{fig:cm7424} and are similar
to those of NGC~2997, except that its cluster population is almost
3\tm\ fainter.  It also has a group of bright, young clusters around (J-K) =
1\fm8 that is separate from the older ones close to 1\fm2.  The populations of
the two branches as a function of \MK\ are more similar than in the case of
NGC~2997.  For the faintest magnitudes below -9\tm, one starts to see old
objects with (J-K)$>$2\tm, which may be individual post TP-AGB stars.  The
CMDs for all galaxies are available in Figs.~\ref{figa:cmall} and
\ref{figa:jkall}.

\begin{table}
 \caption[]{Brightest member and population size of cluster complexes.  The
   fifth brightest absolute magnitude \MK\ is listed using isophotal
   magnitudes, M$_{Ki}$, and aperture estimates, M$_{Ka}$. The number of
   non-stellar sources N$_{12}$ with \MK$<$-12\tm\ is given with `$>$'
   indicating lower limits.  Exponent $\alpha_x$, and number of sources N$_x$
   of a power-law fit to the luminosity function for young and old cluster
   complexes are listed using the subscripts {\it y} and {\it o},
   respectively. }
 \label{tbl:clf}
 \begin{tabular}{lccrcrcr}
\hline\hline
  Galaxy & \multicolumn{1}{c}{M$_{Ki}$} & \multicolumn{1}{c}{M$_{Ka}$} &
  \multicolumn{1}{c}{N$_{12}$} &
  \multicolumn{1}{c}{$\alpha_y$} & \multicolumn{1}{c}{N$_y$} &
  \multicolumn{1}{c}{$\alpha_o$} & \multicolumn{1}{c}{N$_o$}  \\ \hline
    NGC~157  & -17.8 & -14.6 &     453 & 1.62 & 107 & 2.38 &  369 \\
    NGC~1232 & -16.5 & -14.6 &     354 & 2.37 & 142 & 2.54 &  768 \\
    NGC~1300 & -15.9 & -14.8 &     214 & 1.57 &  97 & 2.00 &  193 \\
    NGC~1365 & -18.7 & -15.1 &     705 & 1.97 & 215 & 2.15 &  942 \\
    NGC~1566 & -17.8 & -15.3 &  $>$702 & 1.75 & 339 & 2.62 &  169 \\
    NGC~2997 & -17.6 & -15.2 &    1216 & 2.29 & 189 & 2.49 & 1396 \\
    NGC~4030 & -19.0 & -15.4 &  $>$791 & 2.25 &  45 & 2.31 &  268 \\
    NGC~4321 & -17.4 & -15.6 & $>$1066 & 1.71 & 120 & 2.44 &  463 \\
    NGC~5247 & -17.6 & -15.1 &     974 & 2.19 & 279 & 2.72 &  599 \\
    NGC~7424 & -14.1 & -12.6 &      40 & 1.76 & 512 & 2.38 &  520 \\
  \hline   
\end{tabular}
\end{table}

\section{Luminosity function of clusters}
\label{sec:clf}
Two basic, non-parametric estimates of the luminosity function for the
population of cluster complexes in the galaxies are the brightest complex and
the numbers of sources above a given magnitude limit.  The fifth brightest
non-stellar source was used as an estimator to reduce the influence of
background galaxies since fewer than five of these galaxies are expected for
\Ks$<$16\fm5 according to {\it GalaxyCount}.  Many of the brightest sources
have sizes exceeding the 1\arcsec\ aperture used for the standard photometry.
Although aperture magnitudes still are the more robust indicator of the
luminosity of compact sources, isophotal magnitudes may give a more realistic
estimate of the total flux coming from the largest star-forming complexes.
The fifth brightest complex for each galaxy is listed in Table~\ref{tbl:clf}
as M$_{Ki}$ and M$_{Ka}$ for the isophotal and aperture magnitudes,
respectively.  Nuclear complexes (i.e. $r<20$\arcsec) were excluded as they
are often significantly brighter than those in the disk.  In NGC~1365 and
NGC~4321 in particular, many complexes in the central region were identified
with aperture magnitudes reaching -15\fm5.  Owing to the inclusion of flux
from a larger area, M$_{Ki}$ is 2-3\tm\ brighter than M$_{Ka}$ depending on
the morphology of the individual regions.  As an estimate of the flux
originating from a cluster complex of a single GMC, the aperture magnitude is
accurate because of the close match between the aperture used and the
typical, linear sizes of GMCs.

The size of the cluster population is estimated from the number of non-stellar
sources brighter than a given magnitude.  A limit of \MK=-12\tm\ was used as a
compromise between including some clusters in NGC~7424 and being brighter than
the limiting magnitude for most of the galaxies.  Only three galaxies
(i.e. NGC~1566, NGC~4030, and NGC~4321) did not fully reach -12\tm\ at a
90\% completeness level (see Table~\ref{tbl:gal}).  The number of these
complexes is given as N$_{12}$ in Table~\ref{tbl:clf}.

The analytical form of the cluster luminosity function is normally
approximated by either a power law $ N dL = n L^{-\alpha} dL $ or a Schechter
function, which adds an exponential cut-off at high luminosities.  It is
important to avoid bins affected by incompleteness at the faint end of the
distribution as well as high luminosity outliers (e.g.  background galaxies).
The fixed aperture used for the photometry can lead to underestimates of the
brightest luminosities estimated for large complexes (see Fig.~\ref{fig:iso}),
which can be shifted, for sources with a surface area exceeding the aperture,
to fainter values that depend on their surface brightness.  This uncertainty
in determining magnitudes for the brightest sources made it inappropriate to
use a Schechter function to fit the luminosity function.

\begin{figure*}
  \resizebox{\hsize}{!}{\includegraphics{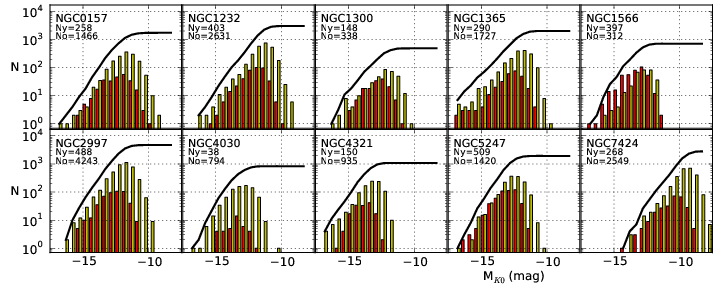}}
  \caption{Numbers of non-stellar sources as a function of their absolute
    magnitude \MKo\ corrected for extinction.  Young objects (0\fm1$<$\Qd) are
    plotted as red columns while older ones (\Qd$<$0\fm1) are shown in yellow.
    Total numbers of young, Ny, and old, No, objects are written below the
    name of the galaxy.  The cumulative cluster luminosity function is shown
    by a full drawn line.}
  \label{fig:mkhist}
\end{figure*}

Even though the K-band is less affected by extinction than visual bands, it is
still important to correct for dust attenuation especially for the younger
complexes, which may have \AK\ reaching almost 1\tm. The distributions of
extinction-corrected absolute magnitudes \MKo\ are shown in
Fig.~\ref{fig:mkhist}, where the cumulative cluster luminosity functions are
indicated by full drawn lines. The distributions of the young objects have
flatter peaks that are shifted to brighter magnitudes than those of
the old ones:  This is mainly due to the higher extinction corrections applied
to the younger population of complexes.  The equal size of the two populations
for the brightest bins is consistent with the mass-independent disruption of
clusters found by \citet{chandar10a} for the Large Magellanic Cloud (LMC).

The histograms provide evidence of neither a significant high luminosity
cut-off nor a steepening of CLF slopes for bright clusters, except possibly
for young complexes in NGC~4030 and NGC~4321. In the magnitude range where
both young and older cluster populations are complete, they have comparable
sizes with two exceptions, namely NGC~1566 with an excess of young clusters
and NGC~4030, which has a deficiency.  Power laws were fitted to the young and
old populations separately with a luminosity range estimated visually from the
\MKo\ distributions.  The exponents and number of sources used are given in
Table~\ref{tbl:clf} and show values of $\alpha$ in the range of 1.6-2.4 for
the younger population while the older ones are slightly steeper.  It is
unclear whether this trend is significant because of the corrections for
extinction, which for the young complexes can reach 1\tm\ in the K-band.

\section{Star formation rate}
\label{sec:sfr}
An estimate of the SFR can be derived from the population of young clusters
assuming that they follow a given evolutionary track. Using a SB99 SSP model
with M$_u = 100$\,\Ms, a Salpeter IMF, and metallicity Z=0.02, a cluster with
a mass of 10$^6$\Ms\ has an absolute magnitude \MK $\approx$ -15\fm6 during
first few Myr.  Clusters become fainter by almost a magnitude when their
massive stars start evolving to the AGB after which they brighten again, as
seen on the evolutionary tracks in Fig.~\ref{fig:cm2997} and
Fig.~\ref{fig:cm7424}.  A CSP model has a smoother variation in color indices
and an increasing luminosity with time depending on its SFR. The SFR of the
galaxies, associated with complexes observed above the limiting magnitude, was
calculated by summing up all sources with 0\fm1$<$\Qd\ assuming that they are
younger than 7\,Myr (as given by the CSP model) and their masses are
proportional to their luminosity with 10$^6$\Ms\ corresponding to
\MKo=-15\fm5.  This value gives a good estimate of the current SFR (i.e.
within the last 7\,Myr) since it is little affected by extinction, in contrast
to estimates relying on UV or H$_{\alpha}$ fluxes which may be absorbed in
dust lanes.  The contribution of clusters below the limiting magnitude can be
estimated by integrating the CLF power-law fit to the young population of
complexes down to a given limit.  Using the exponents listed in
Table~\ref{tbl:clf} and a lower cluster mass of 100\,\Ms, the total SFR$_t$
was computed. These estimates suffer the uncertainty given by the power-law
extrapolation over several orders of magnitude.  In contrast, they are
independent of the different limiting magnitudes and indicate more clearly the
relative star-formation activity in the galaxies.  Both current observed SFR
and extrapolated total SFR$_t$ for the galaxies are listed in
Table~\ref{tbl:sfr}, where we also quote SFR estimates for eight galaxies from
the literature.  Our values are two-to-three times higher for six of the
galaxies, possibly owing to the lower extinction in NIR.  A large amount of
star formation in the central parts of NGC~1566 and NGC~4030 may be the reason
for our lower SFR estimates for these galaxies since crowding and steep
gradients could reduce the completeness of our sample.  The two galaxies also
exhibit the largest differences between bright complexes as measured by
isophotal and aperture magnitudes (see Table~\ref{tbl:clf}).

\begin{figure*}
  \resizebox{\hsize}{!}{\includegraphics{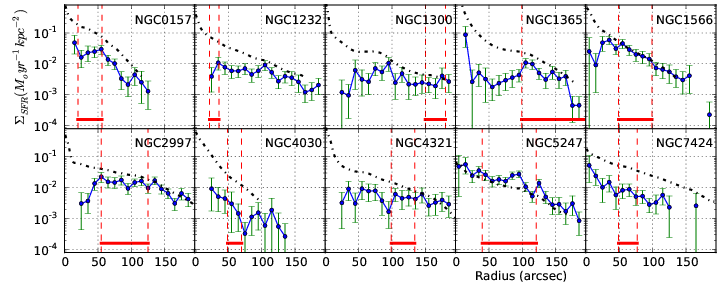}}
  \caption{Radial star formation rates for the galaxies in 10\arcsec\ bins
    with indications of the 10-90 percentile range based on statistical
    fluctuations.  The radial range of the main, symmetric spiral pattern is
    shown by vertical, dashed lines.  The average K-band surface brightness of
    the disk is drawn with a dashed-dotted line in units of kJy kpc$^{-2}$.  }
  \label{fig:sfr}
\end{figure*}

The variation in the SFR per unit area, \SFRa, as a function of Galactocentric
distance is shown in Fig.~\ref{fig:sfr} using the projection parameters listed
in Table~\ref{tbl:gal}.  The errors are dominated by statistical variations
caused by the relatively small number of clusters in each bin.  The mean,
radial \Ks\ surface brightness of the disk is indicated using a zero point of
667 Jy \citep{cohen03}.  We note that \SFRa\ follows the \Ks\ surface
brightness of the disk in most cases with the exception of the bar regions
(see e.g. NGC~1300, NGC~1365, NGC~1566, NGC~2997, and NGC~4321).  The
relatively lower \SFRa\ in the bar regions may be caused by either the radial
redistribution of material (e.g. gas) or a lower star formation efficiency.
Comparing the radial range of the strong, two-armed spiral patterns in the
galaxies with the radial variation in \SFRa, one does not see any significant
change.  This agrees with the results obtained by \citet{elmegreen86} and
\citet{kennicutt98} in the sense that a strong spiral perturbation does not
significantly affect the total \SFRa.

The surface density of young and older non-stellar sources in
10\arcsec\ annular bins is shown in the on-line material
Fig.~\ref{figa:nrhist}.  The radial variations in the two populations follow
each other fairly well in the logarithmic plot indicating that their ratio is
approximately constant.  Only NGC~1566 has a significant excess of young
sources, which is mainly due to its relatively bright limiting magnitudes
restricting the number of older complexes found.

\begin{table*}
 \caption[]{Star formation rates and arm/inter-arm populations in the
   galaxies. The current SFR and the total, extrapolated SFR$_t$ for the
   galaxies are given in \msyr. Inner and outer radii (r$_i$, r$_o$) for the
   region with a symmetric, logarithmic spiral pattern, used for the analysis
   of arm/inter-arm properties, are listed.  The percentile values for the
   Kolmogorov-Smirnov test for arm and inter-arm samples of \Qd\ and
   \MK\ being drawn from the same distribution are shown as P$_Q$ and P$_M$,
   respectively.  The ten percentile values for the \MK\ distributions in arm
   and inter-arms regions are listed as M$_{K10}^a$ and M$_{K10}^i$.  Numbers
   of young cluster complexes, N$_y^a$ and N$_y^i$, in arm and inter-arms
   regions used to estimate the ratio, R$_{a:i}$, of SF in arm to inter-arm
   regions are also listed.  Finally, SFR estimates from the literature are
   shown with references. }
 \label{tbl:sfr}
 \begin{tabular}{lrcrrrrrrrrrc}
\hline\hline
  Galaxy & \multicolumn{1}{c}{SFR}  & \multicolumn{1}{c}{log(SFR$_t$)} &
  \multicolumn{1}{c}{r$_i$} &   \multicolumn{1}{c}{r$_o$} &
  \multicolumn{1}{c}{P$_Q$} &   \multicolumn{1}{c}{P$_M$} &
  \multicolumn{1}{c}{M$_{K10}^a$} & \multicolumn{1}{c}{M$_{K10}^i$} &
  \multicolumn{1}{c}{N$_y^a$} & \multicolumn{1}{c}{N$_y^i$} &
  \multicolumn{1}{c}{R$_{a:i}$} &  \multicolumn{1}{c}{SFR$_{lit}$}
 \\ \hline
    NGC~157  &  3.8 & 2.16 &  20 &  56 & 0.32 & 0.00  & -12.9 & -12.4 &
            43 & 22 & 3.5 &   0.64\tablefootmark{a}, 0.73\tablefootmark{b} \\
    NGC~1232 &  4.2 & 3.61 &  22 &  36 & 0.32 & 0.00  & -11.9 & -11.7 &
            13 &  4 & 4.5 &  2.1\tablefootmark{c}\\
    NGC~1300 &  3.6 & 1.57 & 150 & 180 & 0.20 & 0.93  & -12.8 & -13.0 &
            16 & 10 & 1.1 &  1.4\tablefootmark{c}, 1.7\tablefootmark{d} \\
    NGC~1365 &  5.5 & 2.75 &  99 & 220 & 0.00 & 0.00  & -12.8 & -12.7 &
           129 & 37 & 2.2 &  11.3\tablefootmark{e} \\
    NGC~1566 & 10.9 & 2.86 &  49 &  99 & 0.08 & 0.01  & -13.8 & -13.5 &
           114 & 36 & 5.6 &  9.6\tablefootmark{f} \\
    NGC~2997 & 10.0 & 4.08 &  55 & 125 & 0.37 & 0.00  & -12.9 & -12.4 &
           142 & 63 & 3.1 &  3.7\tablefootmark{g} \\
    NGC~4030 &  1.4 & 3.42 &  49 &  70 & 0.45 & 0.00  & -13.9 & -13.1 &
             2 &  3 &  -- &  11.3\tablefootmark{f} \\
    NGC~4321 &  8.1 & 2.62 &  99 & 134 & 0.97 & 0.24  & -13.8 & -13.5 &
            22 & 22 & 1.2 &  5.2\tablefootmark{b}, 18.5\tablefootmark{f} \\
    NGC~5247 & 13.7 & 4.23 &  40 & 121 & 0.16 & 0.00  & -13.1 & -12.8 &
           180 & 82 & 1.9 &  -- \\
    NGC~7424 &  0.7 & 0.71 &  49 &  77 & 0.70 & 0.20  & -10.2 & -10.1 &
            37 & 33 & 1.7 &  -- \\
  \hline   
\end{tabular}
\tablefoot{References to other SFR estimates corrected to current
  distance scale:
  \tablefoottext{a}{\citet{sempere97}},
  \tablefoottext{b}{\citet{moustakas06}},
  \tablefoottext{c}{\citet{omar05}},
  \tablefoottext{d}{\citet{schulman97}},
  \tablefoottext{e}{\citet{elmegreen09}},
  \tablefoottext{f}{\citet{thilker07}}, and
  \tablefoottext{g}{\citet{kodilkar11}}.
}
\end{table*}

\section{Spatial distribution}
\label{sec:spacial}
The positions of sources within the galaxies were computed from the projection
angles given in Table~\ref{tbl:gal} assuming that they were located in the
plane of the disk.  The shape of the spiral pattern was derived from 1D fast
Fourier transforms (FFT) of the azimuthal, K-band intensity variation in
1\arcsec\ radial bins.  The radial range of the main, two-armed spiral pattern
was estimated by comparing amplitude and phase of the m=2 component, which is
shown in $\theta$-$\ln(r)$ diagrams in Fig.~\ref{figa:alnr}.  The regions
occupied by a symmetric, logarithmic spiral arms are listed in
Table~\ref{tbl:sfr}.

\begin{figure}
  \resizebox{\hsize}{!}{\includegraphics{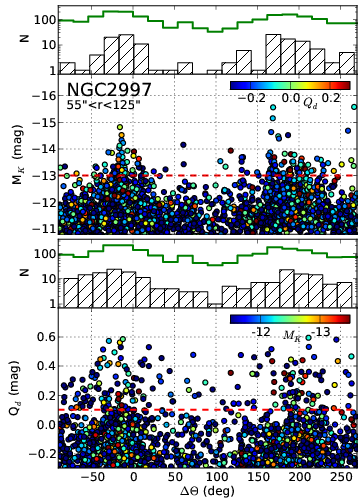}}
  \caption{Absolute magnitude \MK\ and \Qd\ index of non-stellar sources as a
    function of their azimuthal distance from the two-armed spiral pattern,
    where $\Delta\Theta$ = 0\degr\ and 180\degr\ denote the relative \Ks-band
    intensity maxima given by the phase of m=2 FFT component.  The top diagram
    shows \MK\ with colors indicating the \Qd\ index.  The number of clusters
    with \MK\ $>$ -13\tm\ is displayed as a full drawn curve, while the hatched
    histogram shows the brighter clusters.  The low section shows \Qd\ with
    \MK\ used for color coding.  The full drawn histogram indicates older
    clusters with \Qd$<$0\fm1, whereas younger ones are shown by hatched
    bars. }
  \label{fig:mq2997}
\end{figure}

The azimuthal distribution of complexes relative to the two-armed spiral is
shown in Fig.\ref{fig:mq2997} for NGC~2997, considering only the radial range
with a strong, symmetric pattern.  The upper panel displays the absolute
magnitude \MK, which for sources brighter than -13\tm\ shows a strong
concentration in the arm regions.  Fainter clusters have a smoother
distribution as a function of azimuth that also varies in phase with the arms.
A similar behavior can be seen in the lower panel, which shows the \Qd\ index.
Young complexes with 0\fm1$<$\Qd\ are mainly located in the arms, while the
older population has a smaller azimuthal variation.  The concentration in the
arms is stronger for the younger and brighter complexes.  This general trend
is confirmed for all the galaxies (see Fig.~\ref{figa:mqall}) but is more
clearly seen when the spiral perturbation is strong, as in NGC~1365,
NGC~1566, NGC~2997, and NGC~5247.

To investigate general differences in the cluster distributions relative to
the spiral pattern, arm regions were defined as a 90\degr\ azimuthal interval
centered on the phase of the m=2 FFT component, yielding equal areas of arm
and inter-arm regions. This definition allows us to estimate a phase
difference among the individual clusters relative to the spiral pattern. This
is impossible when using the relative azimuthal intensity variation to
distinguish between arm and inter-arm regions.  The contrast between arm and
inter-arm regions varies significantly, where NGC~2997 has a larger amount of
inter-arm complexes than NGC~1365.  The arm and inter-arm populations were
compared to verify whether the two samples could be assumed to be drawn from
the same distribution function but merely be of different sizes, using the
Kolmogorov-Smirnov test and the U-test of Wilcoxon, Man and Witney.  The
distribution function of \Qd\ likely does not to change significantly from the
arm to the inter-arm regions, although the total number of sources varies in
phase with the spiral perturbation.  On the other hand, the distributions of
\MK\ are different in arm and inter-arm regions at a 99\% level of confidence.
The three exceptions (i.e.  NGC~1300, NGC~4321, and NGC~7424) all lack signs
of strong, coherent star formation in the radial regions considered.  The
Kolmogorov-Smirnov statistics are listed in Table~\ref{tbl:sfr} with the
U-tests yielding similar results.  These assumption-free tests are sensitive
to all kinds of differences in the distributions, such as changes in either
the shape or mean.  Comparing the \MK-histograms, the distributions in the
arms for the galaxies with strong spiral perturbations show a shift towards
brighter magnitudes, which is on average 0\fm4 using the ten percentile point
of the distributions as an indicator (see Table~\ref{tbl:sfr}).

We also calculated the SFR in arms relative to inter-arm regions. The ratios
R$_{a:i}$ are listed in Table~\ref{tbl:sfr}, together with the number of young
clusters used to calculate this ratio.  For the galaxies with strong spirals,
the SFR in arms is two-to-five times higher than in the inter-arm regions,
which still have some star-formation activity.

The usage of the m=2 FFT component as a phase reference for the spiral is
acceptable when considering the general arm/inter-arm properties of the
cluster distributions.  The actual arms are imperfectly symmetric and may have
azimuthal shifts of several tens of degrees relative to their average phase.
Different models of star formation in the spiral arms \citep{kim02, kim06,
  dobbs10, gittins04} have distinct predictions regarding the phase
distribution of young clusters relative to the potential minimum of the arms.
To test these predictions, we need to take into account the phase shifts of
the individual arms, any asymmetries in the arm profiles, and color gradients
caused by age and dust variations.  This discussion will be postponed to a
forthcoming paper dealing with the detailed shape of the spiral arms in the
galaxies.

\section{Discussion}
\label{sec:discussion}
The current K-band data are 2-3\tm\ deeper than those presented by
\citet{grosbol08} and have significantly higher resolution, which provide a
much fainter detection limit for cluster complexes.  Only 4 of the 46 galaxies
discussed in the 2008 paper had full JHK photometry.  Although the current
data are of significantly higher quality, it is still of interest to compare
them with the previous results.  The fifth brightest \MK\ aperture magnitudes
agree well, with a spread of 0\fm14, and no significant offset after being
corrected for the different distance scale used.  The largest difference of
0\fm3 was seen for NGC~1566.  Some of the galaxies (e.g. NGC~1365 and
NGC~4321) have bright nuclear clusters that are 1-2\tm more luminous than the
ones in the disk.  The number of non-stellar objects N$_{13}$ brighter than
\MK$<$-13\tm\ was also computed, and corrected for the distance scale.  In
most cases, they were compatible to within 20\%, except for NGC~157, NGC~1566,
and NGC~4030, which all have large, compact star-forming regions.  The higher
resolution of the current data allows us to distinguish smaller complexes and
thereby increase the number of detected sources significantly.  The SFR
estimated by \citet{grosbol08} was based only on the morphology of the
sources, owing to the lack of color information.  In addition, the number of
sources detected was around four times smaller, which explains why the
estimates of the SFR are between two and four times lower than the current
ones.

The distribution of objects in color-color and color-magnitude diagrams is the
primary means of understanding the physical properties of the cluster
complexes.  The (H-K)--(J-H) diagram allows us to distinguish three
populations of sources, namely: a) one with colors around (0\fm4, 0\fm8)
compatible with an old stellar population (i.e. $>$50\,Myr) at low extinction,
b) another with higher (J-K) close to (0\fm8, 1\fm1) indicative of a younger
population with higher extinction, and c) finally objects scattered to higher
(H-K) values, the largest part of which can be interpreted in terms of SB99
models plus extinction from dust. To reproduce the colors of the objects with
the highest (H-K) in this group, we would need to add a substantial amount of
emission from hot dust \citep[see e.g.][]{witt92}.  The relative importance of
these three groups varies significantly from galaxy to galaxy. The total
number of objects in the groups with older sources is sensitive to the
limiting magnitude because clusters get fainter with time. Variations in the
current SFR away from the average will also influence the relative
importance of the old and young groups.

The detailed interpretation of these distributions depends on assumptions such
as the stellar evolutionary models adopted, nebular emission, attenuation by
dust, and the shape of the cluster IMF including its upper mass limit and
slope.  The lack of accurate spectral information for the clusters makes it
impossible to determine the parameters for the IMF in an independent way.  We
adopted the Padova isochrones \citep{marigo08} as they include the late
evolutionary phases (e.g. post TP-AGB), which are important for the NIR
colors.  A main difference between the two sets of cluster tracks shown in
Fig.~\ref{fig:cc2997} is either the inclusion or exclusion of nebular emission
when estimating the NIR colors for SB99 or Padova tracks, respectively.  The
inclusion makes (H-K) significantly redder in the early phases when hot stars
are present.  The mode of star formation also affects the integrated
colors, where a continuous formation in the clusters prolongs the time-span
during which nebular and \Brg\ emission are significant.  Without a good
estimate of the intrinsic colors of the complexes, the appropriate reddening
law can only be selected based on general considerations.  It is reasonable to
assume that the attenuation by dust of the integrated light is described by a
law somewhere between a `screen' and `dusty' model \citep{witt92} since nearby
star forming regions contain significant amounts of dust.  The reddening vector
could even be time dependent if supernovae are able to remove a major fraction
of the dust from the inner regions of the clusters.  This scenario
suggests that the 'dusty' model should be applied to the youngest complexes,
while the older ones may be described by a `screen' model.

The correlation between \Brg-emission and the reddening-corrected Q index for
NGC~2997 (see Fig.~\ref{fig:brg}) clearly indicates that Q can be used to
separate very young complexes (i.e. $\la$7\,Myr) from older ones.  The
detailed relation between Q and cluster age depends on both the model tracks
and reddening vector assumed.

The (J-K)--\MK\ diagrams show a double-branch structure (most clearly
separated for NGC~1566, NGC~2997, and NGC~7424) for bright complexes with
(J-K) $\approx$ 1\fm2 and 1\fm8, respectively.  The main branch around (J-K) =
1\fm2 corresponds well to the location predicted for old clusters with
relatively low extinction.  The model tracks suggest that young clusters have
lower (J-K) values reaching almost 0\fm0 for ages younger than one Myr.  This
region is virtually empty, while the majority of the young complexes with
0\fm1$<$\Qd\ are grouped near (J-K) = 1\fm8 suggesting almost 10\tm\ of visual
extinction.  Since young complexes must migrate to the branch around 1\fm2,
the gap between the two suggests that a rapid reduction in the extinction (i.e.
$\la$10\, Myr) occurs just after \Brg-emission from the complexes ceases.  The
reddening-corrected Q index shows a more continuous distribution, supporting
the suggestion that the separation of the two branched in (J-K) is due to an
expulsion of dust rather than a sudden change in the SFR.

Information on the early phase of the complexes can be obtained by comparing
the location of the young sources with high extinction in the CCD, to the
model tracks.  The shape of this group is mainly elongated along the direction
of the reddening vector.  This is inconsistent with a SB99 SSP model and a
smooth SFR in time, which would yield a much broader distribution
perpendicular to the reddening vector than observed.  Most CSP models would be
feasible because it would take a considerable time, several Myr, before the
clusters would accumulate enough mass to be brighter than the detection
limit. During the earliest cluster-evolutionary-phase the main sequence
lifetime of the more massive ionizing stars is clearly comparable to their
contraction time, which would mimic a continuous star formation.  Later on,
when those stars evolved, Padova SSP models represent well the CCDs.  Also
models with low nebular emission would be acceptable in which case the
estimated visual extinctions could reach \AV\,=\,12\tm.  It is not possible to
distinguish between these scenarios based on broad band photometry only.

For comparison, we used the \citet{pessev06} sample of Small Magellanic Cloud
(SMC) and LMC clusters with integrated colors obtained from 2MASS. Those
clusters also show a distribution similar to those in our sample of galaxies,
indicating that our sample also includes intermediate and old globular-like
objects.  Even the peculiar colors, such as those observed in several clusters
in the galaxies NGC~7424 and NGC~1365 with (J-K)$<$0\fm1 and (H-K)$>$0\fm5
(lower right part of CCD), can also be found in the SMC-LMC clusters, such as
NGC~458 (200\,Myr), NGC~2153 (1.3\,Gyr), and NGC~2172 (40\,Myr).

The CLFs can be compared to those derived by \citet{larsen02} from HST UBV
photometry.  Their mean exponent for the power-law function is
$\alpha=2.26\pm0.16$, while our mean exponents for old and young clusters are
$\alpha_o=2.40\pm0.21$ and $\alpha_y=1.95\pm0.30$, respectively.  As can be
seen, the mean exponent derived from the HST observations is more similar to
that for our old cluster mean exponent.  This result is consistent with the
younger objects having higher extinctions, which would probably hide many of
them in UV and optical bands analyzed by \citet{larsen02}.  We do not find any
evidence of the high luminosity cut-off reported by \citet{gieles06}.  Again
here, the usage of NIR photometry allows us to obtain a more complete census
of the youngest and brightest clusters which may be obscured by dust lanes at
shorter wavelengths.  Our results are consistent with \citet{chandar10,
  chandar10a} who estimated an exponent of 1.80 for the Magellanic Clouds and
found no evidence of a high-luminosity cut-off.  Since NIR colors provide
rather limited amount of age information for older sources, we were unable to
determine whether the properties of the CLF depend on the cluster age.

The distributions of the \MK\ and \Qd\ of sources as a function of the
azimuthal phase relative to the spiral arms show a characteristic variation in
the regions of strong, symmetric perturbations.  The total number density of
complexes oscillates in phase with the K-band surface brightness variations.
There is no statistical evidence for a change in the \Qd\ distribution
function from the arm to the inter-arm regions.  There is clear evidence that
the \MK\ distribution in the arms differs from that in-between.  The
brightest complexes are preferentially found in arms.  The existence of
non-linear compressions or shocks in the gas could alter the mass spectrum and
allow more massive clusters to be formed in the arm regions.

\section{Conclusion}

Our analysis of NIR photometry and location of cluster complexes relative to
major arms in 10 nearby, grand-design, spiral galaxies had led to the following
conclusions:

\begin{itemize}
\itemsep=0pt

\item All complexes younger than 7\,Myr have significant extinction in
  the range \AV=3-7\tm, while older ones typically have \AV$<$1\tm.

\item The (J-K)--\MK\ diagrams show two well-separated branches,
  which are indicative of a rapid expulsion of dust around an age of 7\,Myr.

\item The CLFs are well-fitted by power laws with an exponent $-1.95\pm0.3$
  for clusters younger than 7\,Myr and $2.40\pm0.2$ for older clusters. The
  distributions do not show evidence of a cut-off at their high luminosity
  end.

 \item Cluster complexes formed in strong spiral arms have a CLF that is
   systematically brighter in \MK\ by at least 0\fm4 than those in the
   inter-arm regions.

\item The azimuthal average of \SFRa\ is insignificantly affected by
  strong spiral perturbations, which help to concentrate star formation in
  the arm regions.  Outside the grand-design spiral pattern, star formation is
  more uniformly distributed in azimuth.

\item The SFR in the arm regions of galaxies with strong spiral
  perturbations is 2-5 times higher than in the inter-arm regions.
  
\end{itemize}

The NIR observations of cluster complexes in grand-design galaxies clearly
reveal a population of very young clusters highly attenuated by dust.  These
clusters are difficult to access in visual bands but are essential for
understanding the early evolution of clusters.  The indication of a shift in
their CLF, towards brighter clusters, in the arms points to a change in the
physical conditions.  This suggests a scenario where star formation is
enhanced by spiral perturbations (e.g. by increased shear or gas densities, or
shocks) but in inter-arm regions is due to stochastic processes.
A full analysis of the formation of clusters in the arms of grand-design
spirals should include dynamical models for the flow of material.

\begin{acknowledgements}
  The ESO-MIDAS system and {\it SExtreator} were used during the reduction and
  analysis of the data.  We would also like to thank the anonymous referee for
  comments that helped us to improve the presentation. This research has made
  use of the NASA/IPAC Extragalactic Database (NED), which is operated by the
  Jet Propulsion Laboratory, California Institute of Technology, under
  contract with the National Aeronautics and Space Administration.
\end{acknowledgements}
\bibliographystyle{aa}
\bibliography{AstronRef}

\begin{thebibliography}{71}
\expandafter\ifx\csname natexlab\endcsname\relax\def\natexlab#1{#1}\fi

\bibitem[{Bastian \& Goodwin(2006)}]{bastian06}
Bastian, N. \& Goodwin, S.~P. 2006, MNRAS, 369, L9

\bibitem[{Becker {et~al.}(1988)Becker, Mebold, Reif, \& {van
  Woerden}}]{becker88}
Becker, R., Mebold, U., Reif, K., \& {van Woerden}, H. 1988, A\&A, 203, 21

\bibitem[{Becker \& Fenkart(1970)}]{becker70}
Becker, W. \& Fenkart, R. 1970, in IAU Symp., Vol.~38, The Spiral Structure of
  our Galaxy, ed. W.~Becker \& G.~Contopoulos, 205

\bibitem[{Bertin \& Arnouts(1996)}]{bertin96}
Bertin, E. \& Arnouts, S. 1996, A\&AS, 117, 393

\bibitem[{Bertin {et~al.}(1989)Bertin, Lin, Lowe, \& Thurstans}]{bertin89}
Bertin, G., Lin, C.~C., Lowe, S.~A., \& Thurstans, R.~P. 1989, ApJ, 338, 78

\bibitem[{Bigiel {et~al.}(2008)Bigiel, Leroy, Walter, Brinks, de~Blok, Madore,
  \& Thornley}]{bigiel08}
Bigiel, F., Leroy, A., Walter, F., {et~al.} 2008, AJ, 136, 2846

\bibitem[{Chandar {et~al.}(2010{\natexlab{a}})Chandar, Fall, \&
  Whitmore}]{chandar10a}
Chandar, R., Fall, S.~M., \& Whitmore, B.~C. 2010{\natexlab{a}}, ApJ, 711, 1263

\bibitem[{Chandar {et~al.}(2010{\natexlab{b}})Chandar, Whitmore, \&
  Fall}]{chandar10}
Chandar, R., Whitmore, B.~C., \& Fall, S.~M. 2010{\natexlab{b}}, ApJ, 713, 1343

\bibitem[{Cohen {et~al.}(2003)Cohen, Wheaton, \& Megeath}]{cohen03}
Cohen, M., Wheaton, W.~A., \& Megeath, S.~T. 2003, AJ, 126, 1090

\bibitem[{{de Vaucouleurs} {et~al.}(1991){de Vaucouleurs}, {de Vaucouleurs},
  Cowien, {et~al.}}]{rc3}
{de Vaucouleurs}, G., {de Vaucouleurs}, A., Cowien, H., {et~al.} 1991, Third
  reference catalogue of bright galaxies (New York: Springer)

\bibitem[{Dobbs {et~al.}(2010)Dobbs, Theis, Pringle, \& Bate}]{dobbs10}
Dobbs, C.~L., Theis, C., Pringle, J.~E., \& Bate, M.~R. 2010, MNRAS, 403, 625

\bibitem[{Ellis \& Bland-Hawthorn(2006)}]{ellis06}
Ellis, S.~C. \& Bland-Hawthorn, J. 2006, AAO Newsletter, 110, 16

\bibitem[{Elmegreen \& Elmegreen(1986)}]{elmegreen86}
Elmegreen, B.~G. \& Elmegreen, D.~M. 1986, ApJ, 311, 554

\bibitem[{Elmegreen {et~al.}(2009)Elmegreen, Galliano, \& Alloin}]{elmegreen09}
Elmegreen, B.~G., Galliano, E., \& Alloin, D. 2009, ApJ, 703, 1297

\bibitem[{Foyle {et~al.}(2011)Foyle, Rix, Dobbs, Leroy, \& Walter}]{foyle11}
Foyle, K., Rix, H.-W., Dobbs, C.~L., Leroy, A.~K., \& Walter, F. 2011, ApJ,
  735, 101

\bibitem[{Foyle {et~al.}(2010)Foyle, Rix, \& Zibetti}]{foyle10}
Foyle, K., Rix, H.-W., \& Zibetti, S. 2010, MNRAS, 407, 163

\bibitem[{Fujii {et~al.}(2011)Fujii, Baba, Saitoh, Makino, Kokubo, \&
  Wada}]{fujii11}
Fujii, M.~S., Baba, J., Saitoh, T.~R., {et~al.} 2011, ApJ, 730, 109

\bibitem[{Ganda {et~al.}(2006)Ganda, Falc{\'o}n-Barroso, Peletier, Cappellari,
  Emsellem, McDermid, {et~al.}}]{ganda06}
Ganda, K., Falc{\'o}n-Barroso, J., Peletier, R.~F., {et~al.} 2006, MNRAS, 367,
  46

\bibitem[{Gieles {et~al.}(2006)Gieles, Larsen, Bastian, \& Stein}]{gieles06}
Gieles, M., Larsen, S.~S., Bastian, N., \& Stein, I.~T. 2006, A\&A, 450, 129

\bibitem[{Gittins \& Clarke(2004)}]{gittins04}
Gittins, D.~M. \& Clarke, C.~J. 2004, MNRAS, 349, 909

\bibitem[{Goodwin \& Bastian(2006)}]{goodwin06}
Goodwin, S.~P. \& Bastian, N. 2006, MNRAS, 373, 752

\bibitem[{Grosb{\o}l \& Dottori(2008)}]{grosbol08}
Grosb{\o}l, P. \& Dottori, H. 2008, A\&A, 490, 87

\bibitem[{Grosb{\o}l \& Dottori(2009)}]{grosbol09}
---. 2009, A\&A, 499, L21

\bibitem[{Grosb{\o}l {et~al.}(2006)Grosb{\o}l, Dottori, \& Gredel}]{grosbol06}
Grosb{\o}l, P., Dottori, H., \& Gredel, R. 2006, A\&A, 453, L25

\bibitem[{Grosb{\o}l \& Patsis(1998)}]{gp98}
Grosb{\o}l, P. \& Patsis, P.~A. 1998, A\&A, 336, 840

\bibitem[{Grosb{\o}l {et~al.}(2004)Grosb{\o}l, Patsis, \& Pompei}]{grosbol04}
Grosb{\o}l, P., Patsis, P.~A., \& Pompei, E. 2004, A\&A, 423, 849

\bibitem[{Hess {et~al.}(2009)Hess, Pisano, Wilcots, \& Chengalur}]{hess09}
Hess, K.~M., Pisano, D.~J., Wilcots, E.~M., \& Chengalur, J.~N. 2009, ApJ, 699,
  76

\bibitem[{Hodgkin {et~al.}(2009)Hodgkin, Irwin, Hewett, \& Warren}]{hodgkin09}
Hodgkin, S.~T., Irwin, M.~J., Hewett, P.~C., \& Warren, S.~J. 2009, MNRAS, 394,
  675

\bibitem[{Indebetouw {et~al.}(2005)Indebetouw, Mathis, Babler, Meade, Watson,
  Whitney, {et~al.}}]{indebetouw05}
Indebetouw, R., Mathis, J.~S., Babler, B.~L., {et~al.} 2005, ApJ, 619, 931

\bibitem[{Israel {et~al.}(1998)Israel, van~der Werf, Hawarden, \&
  Aspin}]{israel98}
Israel, F.~P., van~der Werf, P.~P., Hawarden, T.~G., \& Aspin, C. 1998, A\&A,
  336, 433

\bibitem[{Kennicutt(1989)}]{kennicutt89}
Kennicutt, R.~C. 1989, ApJ, 344, 685

\bibitem[{Kennicutt(1998)}]{kennicutt98}
---. 1998, ARA\&A, 36, 189

\bibitem[{Kim \& Ostriker(2002)}]{kim02}
Kim, W.-T. \& Ostriker, E.~C. 2002, ApJ, 570, 132

\bibitem[{Kim \& Ostriker(2006)}]{kim06}
---. 2006, ApJ, 646, 213

\bibitem[{Knapen {et~al.}(1993)Knapen, Cepa, Beckman, {Soledad del Rio}, \&
  Pedlar}]{knapen93}
Knapen, J.~H., Cepa, J., Beckman, J.~E., {Soledad del Rio}, M., \& Pedlar, A.
  1993, ApJ, 416, 563

\bibitem[{Kodilkar {et~al.}(2011)Kodilkar, Kantharia, \&
  Ananthakrishnan}]{kodilkar11}
Kodilkar, J., Kantharia, N.~G., \& Ananthakrishnan, S. 2011, MNRAS, 416, 522

\bibitem[{Kuno {et~al.}(2007)Kuno, Sato, Nakanishi, Hirota, Tosaki, Shioya,
  {et~al.}}]{kuno07}
Kuno, N., Sato, N., Nakanishi, H., {et~al.} 2007, PASJ, 59, 117

\bibitem[{Lada \& Lada(2003)}]{lada03}
Lada, C.~J. \& Lada, E.~A. 2003, ARA\&A, 41, 57

\bibitem[{Larsen(2002)}]{larsen02}
Larsen, S.~S. 2002, AJ, 124, 1393

\bibitem[{Larsen(2009)}]{larsen09}
---. 2009, A\&A, 494, 539

\bibitem[{Larsen \& Richtler(1999)}]{larsen99}
Larsen, S.~S. \& Richtler, T. 1999, A\&A, 345, 59

\bibitem[{Leitherer {et~al.}(1999)Leitherer, Schaerer, Goldader, {Gonz{\'a}lez
  Delgado}, Robert, {et~al.}}]{leitherer99}
Leitherer, C., Schaerer, D., Goldader, J.~D., {et~al.} 1999, ApJS, 123, 3

\bibitem[{Lin \& Shu(1964)}]{lin64}
Lin, C.~C. \& Shu, F.~H. 1964, ApJ, 140, 646

\bibitem[{Lindblad {et~al.}(1997)Lindblad, Kristen, J{\"o}rs{\"a}ter, \&
  H{\"o}gbom}]{lindblad97}
Lindblad, P. A.~B., Kristen, H., J{\"o}rs{\"a}ter, S., \& H{\"o}gbom, J. 1997,
  A\&A, 317, 36

\bibitem[{Marigo {et~al.}(2008)Marigo, Girardi, Bressan, Groenewegen, Silva, \&
  Granato}]{marigo08}
Marigo, P., Girardi, L., Bressan, A., {et~al.} 2008, A\&A, 482, 883

\bibitem[{Moustakas \& Kennicutt(2006)}]{moustakas06}
Moustakas, J. \& Kennicutt, R.~C. 2006, ApJ, 651, 155

\bibitem[{Omar \& Dwarakanath(2005)}]{omar05}
Omar, A. \& Dwarakanath, K.~S. 2005, JApA, 26, 89

\bibitem[{Patsis {et~al.}(2010)Patsis, Kalapotharakos, \&
  Grosb{\o}l}]{patsis10}
Patsis, P.~A., Kalapotharakos, C., \& Grosb{\o}l, P. 2010, MNRAS, 408, 22

\bibitem[{Pence {et~al.}(1990)Pence, Taylor, \& Atherton}]{pence90}
Pence, W.~D., Taylor, K., \& Atherton, P. 1990, ApJ, 357, 415

\bibitem[{Pessev {et~al.}(2006)Pessev, Goudprooij, Puzia, \&
  Chandar}]{pessev06}
Pessev, P.~M., Goudprooij, P., Puzia, T.~H., \& Chandar, R. 2006, AJ, 132, 781

\bibitem[{Rix \& Rieke(1993)}]{rix93a}
Rix, H.-W. \& Rieke, M.~J. 1993, ApJ, 418, 123

\bibitem[{Roberts(1969)}]{roberts69a}
Roberts, W.~W. 1969, ApJ, 158, 123

\bibitem[{Salpeter(1955)}]{salpeter55}
Salpeter, E.~E. 1955, ApJ, 121, 161

\bibitem[{Sandage \& Tammann(1981)}]{sandage81}
Sandage, A. \& Tammann, G.~A. 1981, A {R}evised {S}hapley-{A}mes {C}atalog of
  {B}right {G}alaxies, Carnegie Inst. of Wash. Publ. No. 635 (Washington:
  Carnegie Inst.)

\bibitem[{Schechter(1976)}]{schechter76}
Schechter, P. 1976, ApJ, 203, 297

\bibitem[{Scheepmaker {et~al.}(2007)Scheepmaker, Haas, Gieles, Bastian, Larsen,
  \& Lamers}]{scheepmaker07}
Scheepmaker, R.~A., Haas, M.~R., Gieles, M., {et~al.} 2007, A\&A, 469, 925

\bibitem[{Scheepmaker {et~al.}(2009)Scheepmaker, Lamers, Anders, \&
  Larsen}]{scheepmaker09}
Scheepmaker, R.~A., Lamers, H. J. G. L.~M., Anders, P., \& Larsen, S.~S. 2009,
  A\&A, 494, 81

\bibitem[{Schlegel {et~al.}(1998)Schlegel, Finbeiner, \& Davis}]{schlegel98}
Schlegel, D.~J., Finbeiner, D.~P., \& Davis, A. 1998, ApJ, 500, 525

\bibitem[{Schmidt(1959)}]{schmidt59}
Schmidt, M. 1959, ApJ, 129, 243

\bibitem[{Schulman {et~al.}(1997)Schulman, Ockels, \& Knezek}]{schulman97}
Schulman, E., Ockels, F., \& Knezek, P.~M. 1997, BAAS, 29, 1332

\bibitem[{Seigar \& James(1998)}]{seigar98}
Seigar, M.~S. \& James, P.~A. 1998, MNRAS, 299, 672

\bibitem[{Seigar \& James(2002)}]{seigar02b}
---. 2002, MNRAS, 337, 1113

\bibitem[{Sellwood(2010)}]{sellwood10}
Sellwood, J.~A. 2010, MNRAS, 409, 145

\bibitem[{Sellwood(2011)}]{sellwood11}
---. 2011, MNRAS, 410, 1637

\bibitem[{Sempere \& Rozas(1997)}]{sempere97}
Sempere, M.~J. \& Rozas, M. 1997, A\&A, 317, 405

\bibitem[{Skrutskie {et~al.}(2006)Skrutskie, Cutri, Stiening, Weinberg,
  Schneider, Carpenter, {et~al.}}]{2mass}
Skrutskie, M.~F., Cutri, R.~M., Stiening, R., {et~al.} 2006, AJ, 131, 1163

\bibitem[{Thilker {et~al.}(2007)Thilker, Bianchi, Meurer, de~Paz, Boissier,
  Madore, {et~al.}}]{thilker07}
Thilker, D.~A., Bianchi, L., Meurer, G., {et~al.} 2007, ApJS, 173, 538

\bibitem[{{van Zee} \& Bryant(1999)}]{vanzee99}
{van Zee}, L. \& Bryant, J. 1999, AJ, 118, 2172

\bibitem[{Wada {et~al.}(2011)Wada, Baba, \& Saitoh}]{wada11}
Wada, K., Baba, J., \& Saitoh, T.~R. 2011, ApJ, 735, 1

\bibitem[{Witt {et~al.}(1992)Witt, Thronson, \& {Capuano, Jr.}}]{witt92}
Witt, A.~N., Thronson, H.~A., \& {Capuano, Jr.}, J.~M. 1992, ApJ, 393, 611

\bibitem[{Zurita {et~al.}(2002)Zurita, Beckman, Rozas, \& Ryder}]{zurita02}
Zurita, A., Beckman, J.~E., Rozas, M., \& Ryder, S. 2002, A\&A, 386, 801

\end{thebibliography}

\Online
\begin{appendix}
\section{Morphology of galaxies and distribution of cluster complexes}
\label{app:gal}
\begin{figure*}
  \resizebox{0.95\hsize}{!}{\includegraphics{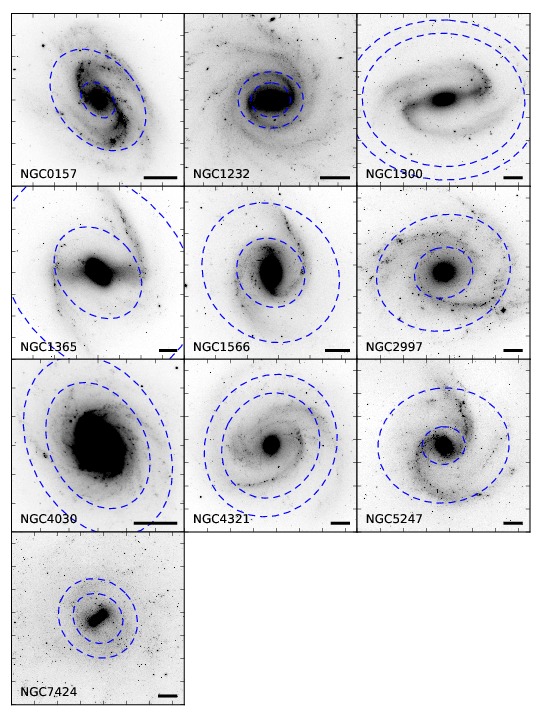}}
  \caption{\Ks-maps of all the galaxies in the sample in negative
    representation. Both intensity and angular scale have been varied to
    help enhance the spiral structure in the galaxies. The two dashed ellipses
    indicate the radial region of the main, symmetric spiral pattern.  All
    images are orientated with north to the top and east to the left with a
    scale indicated by the 30\arcsec\ bar in the lower right corner. }
  \label{figa:gall}
\end{figure*}

\begin{figure*}
  \resizebox{0.92\hsize}{!}{\includegraphics{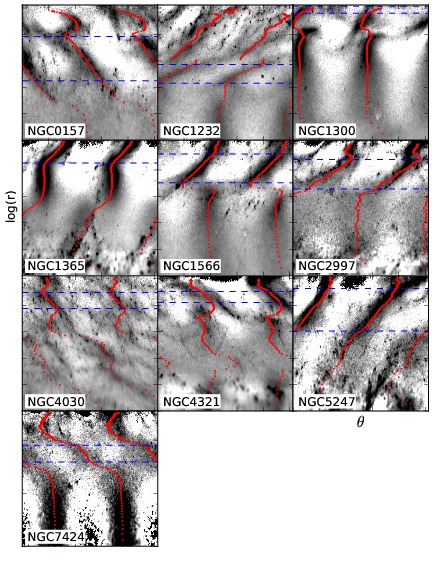}}
  \caption{$\Theta$-$\ln(r)$ maps of the galaxies showing the variation
    relative to the radial \Ks\ intensity profile using the projection angles
    listed in Table~\ref{tbl:gal}. A negative gray scale is used from 0.5
    (white) to 1.5 (black).  A spherical bulge component was fitted and
    removed before de-projecting the galaxies.  The azimuthal angle spans
    2$\pi$ starting from the position angle of the galaxy while the radius
    begins at 5\arcsec. The two dashed lines indicate the radial range of the
    symmetric, logarithmic spiral pattern identical to the two ellipses in
    Fig.~\ref{figa:gall}.  The phase of the m=2 Fourier component is given by
    dots.  }
  \label{figa:alnr}
\end{figure*}

The K-band images of the ten grand-design, spiral galaxies in the sample are
shown in Fig~\ref{figa:gall} as direct images, while $\theta$-$\ln(r)$
representations are given in Fig~\ref{figa:alnr}.  A short description of
their spiral structure is given below based on the Fourier analysis of the
azimuthal intensity variation in their de-projected disks.

\begin{description}
\item[NGC~157:] The galaxy has a weak oval distortion at its center, which
  reaches to a radius of around 10\arcsec\ and has a relative amplitude of
  $a_2 \approx$~2\%. There is a transition zone to 20\arcsec, where a
  two-armed, logarithmic spiral with a pitch angle of 33.4\degr\ starts.  Its
  relative amplitude peaks around 40\% in the range 30\arcsec$<r<$50\arcsec.
  The main m=2 pattern terminates close to 60\arcsec, outside which a weaker,
  tighter pattern exists.

\item[NGC~1232:] It has been argued that this galaxy has a morphology similar
  to that of our Milky Way \citep{becker70}.  It has a bar out to a radius of
  16\arcsec\ with $a_2$ = 20\%.  A tight, symmetric, two-armed spiral with
  ($a_2$, $i$) = (0.2, 14.7\degr) emerges from the end of the bar. Its
  symmetric part terminates at $r=36\arcsec$, after which the northern arm
  splits as the southern arm continues.  A more open pattern can be traced to
  120\arcsec.

\item[NGC~1300:] There are indications of a weak central bar with $a_2$ = 10\%
  within 8\arcsec.  The main bar reaches an amplitude $a_2$ = 55\% around
  64\arcsec, where it widens into ansae.  A tightly wound set of arms emerges
  from the end of the bar close to 95\arcsec, while a faint spiral pattern can
  be traced outside 150\arcsec with ($a_2$, $i$) = (0.2, 13.7\degr).  A
  dynamical study by \citet{patsis10} suggests that the inner arms are
  supported by stars on chaotic orbits that escape through the Lagrangian
  points at the end of the bar.

\item[NGC~1365:] A spiral structure occupies the central parts, while the bar
  is seen in the range 40-92\arcsec\ with an amplitude $a_2$ up to 73\%.  The
  main spiral starts at the ends of the bar and follows a logarithmic spiral
  with $i$ = 33.1\degr\ from 99\arcsec\ to the edge of the frame around
  220\arcsec.  The amplitude of its m=2 component exceeds unity in the outer
  parts owing to its strongly peaked, azimuthal shape.

\item[NGC~1566:] The central part hosts a weak oval distortion with a
  position angle of \~30\degr.  Two symmetric regions in the radial range of
  15-35\arcsec\ and N-S direction show strong star formation activity.  The
  arms connect to these regions but the logarithmic spiral first starts at $r$
  = 49\arcsec\ with ($a_2$, $i$) = (0.7, 27.1\degr).  Outside 99\arcsec, the
  spiral pattern changes shape.

\item[NGC~2997:] Besides a ring of star forming regions at the center, a bar
  with $a_2$ = 10\% ends around 45\arcsec, where the main spiral emerges with
  ($a_2$, $i$) = (0.4, 21.2\degr).  The regular logarithmic spiral pattern
  terminates close to 125\arcsec, where the southern arm has an abrupt turn.

\item[NGC~4030:] The inner part has a complex structure with multiple spiral
  arms.  The main symmetric, two-armed spiral pattern is located in the region
  49-70\arcsec\ with ($a_2$, $i$) = (0.3, 26.5\degr).

\item[NGC~4321:] The inner part of NGC~4321 is occupied by a central bar
  inside 13\arcsec, a main bar to around 67\arcsec and several ring/spiral
  structures.  A set of prominent spiral arms starts at the end of the bar but
  these are asymmetric (with a much stronger southern arm) and tightly wound,
  in a similar way to those of NGC~1300.  Close to 99\arcsec, a logarithmic
  spiral can be traced to 134\arcsec\ with the parameters ($a_2$, $i$) = (0.5,
  29.1\degr).

\item[NGC~5247:] The spiral pattern of this galaxy, as for that of NGC~1566,
  is a prototype clean, open m=2 logarithmic spiral.  There is evidence of
  both a central bar within 5\arcsec\ and a weak spiral pattern reaching
  8\arcsec\ that has the opposite winding of the main spiral.  After a
  transition zone, the main spiral pattern follows a logarithmic spiral with
  ($a_2$, $i$) = (0.6, 34.4\degr) in the range 40\arcsec\ to 121\arcsec. There
  is a small phase shift around 100\arcsec. The azimuthal profile is peaked
  with significant amplitudes of higher-order even harmonics.

\item[NGC~7424:] This galaxy is the closest in the sample with a distance of
  9.5~Mpc.  With a linear resolution of 20~pc, most sources are likely to be
  associated with individual clusters.  There are no traces of either
  significant dust lanes or large star-forming regions.  Its bar extends to
  38\arcsec, where a tight spiral emerges at an angle of almost 90\degr.  The
  main set of symmetric arms starts around 49\arcsec\ with ($a_2$, $i$) =
  (0.3, 12.3\degr) and terminates at 77\arcsec.  Another tight spiral can be
  traced out to 90\arcsec\ with ($a_2$, $i$) = (0.1, 9.0\degr).  The spirals
  have a smooth, azimuthal shape with little power at higher frequencies than
  m=2.

\end{description}

\begin{figure*}
  \resizebox{0.95\hsize}{!}{\includegraphics{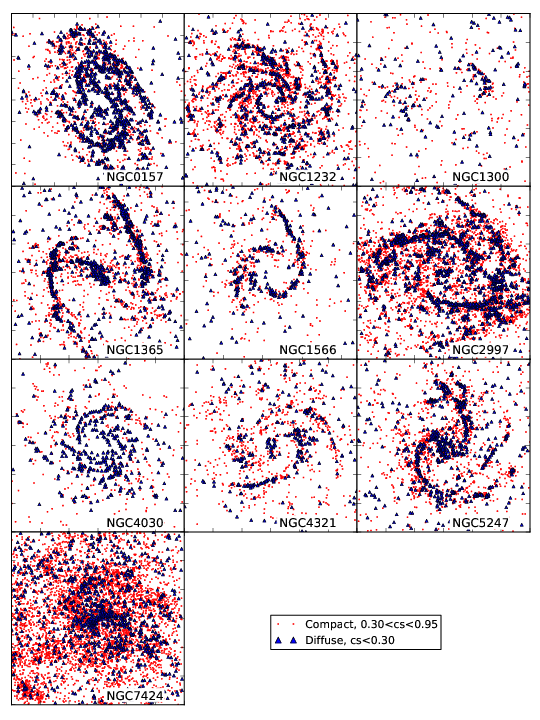}}
  \caption{Positions of all non-stellar sources with photometric errors
    $<$0\fm3 identified on the \Ks-maps.  Compact sources ($0.3<cs<0.95$) are
    indicated by points (red) while diffuse ones ($cs<0.3$) are plotted as
    triangles (blue).  The scale of the images corresponds to that of
    Fig.\ref{figa:gall}.}
  \label{figa:cxy}
\end{figure*}

The positions of all non-stellar sources with photometric errors $<$0\fm3 are
shown in Fig.~\ref{figa:cxy}, where compact complexes ($0.3<cs<0.95$) are
plotted as red points and diffuse ones ($cs<0.3$) as blue triangles.  The
diffuse objects outline the main spiral structure more clearly than the
compact ones which are more uniformly distributed.  This is mainly due to a
larger fraction of compact sources having \Qd$<$0\fm1 and therefore being
older on average than the diffuse sources.

\section{Color and magnitude diagrams}
\label{app:ccm}

\begin{figure*}
  \resizebox{0.95\hsize}{!}{\includegraphics{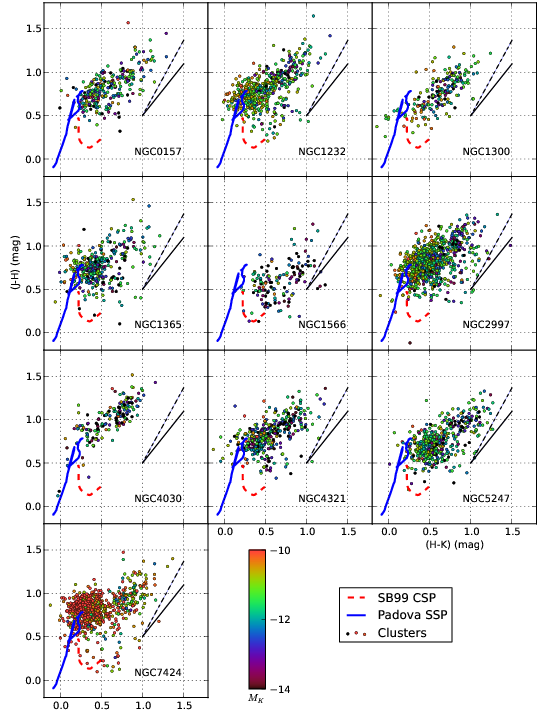}}
  \caption{(H-K)--(J-H) diagrams for non-stellar sources with errors
    $<$0\fm05. The blue full-drawn line show an evolutionary track of a Padova
    SSP model, while the red dashed line indicates a corresponding SB99 CSP
    model (for details see Fig.~\ref{fig:cc2997}). The two black lines show
    the directions of the reddening vectors for `screen' and `dusty' models
    (see Fig.~\ref{fig:cc2997}). Colors indicate the \MK\ of the clusters. }
  \label{figa:ccall}
\end{figure*}

The (H-K)--(J-H) diagrams of non-stellar sources in the galaxies are given in
Fig.~\ref{figa:ccall}, where colors represent their \MK.  For reference, the
`screen' and `dusty' reddening vectors and cluster evolutionary tracks are
indicated in Fig.~\ref{fig:cc2997}.  They have the same general features as
those described in Sect.~\ref{sec:ccd} but there is a significant variation in
their relative importance. To a large extent, this is due to differences in
the limiting magnitude since the `older' clump is dominated by faint sources.
The average location of the high extinction clump varies mainly in (J-H),
which may be explained by slight changes in the effective extinction law
applicable to the complexes.

\begin{figure*}
  \resizebox{0.98\hsize}{!}{\includegraphics{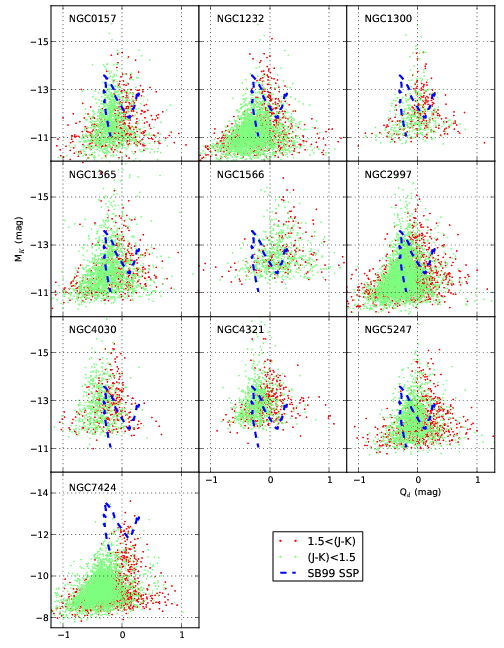}}
  \caption{Absolute \Ks-band magnitude \MK\ of all non-stellar sources with
    errors $<$0\fm3 as a function of their \Qd\ index. The dashed line shows a
    SB99 SSP model with a mass of $10^5$\,\Ms. The color code indicates
    whether the (J-K) index is either above (red/dark) or below 1\fm5
    (green/light).}
  \label{figa:cmall}
\end{figure*}

The \Qd-\MK\ diagrams are presented in Fig.\ref{figa:cmall} with the standard
SB99 SSP model for a cluster of $10^5$\,\Ms.  All the galaxies except for
NGC~7424 have complexes reaching absolute magnitudes \MK\ close to -15\tm.
NGC~1365 and NGC~4321 have several brighter sources, which all are located in
their central regions.  The brightest complexes are found near \Qd=0\fm0 but
would be shifted to slightly higher luminosities if extinction corrections were
applied.  The decreasing number of sources moving from \Qd=0\fm0 to higher
values is caused by the youngest objects, with higher \Qd, still forming stars
and increasing their luminosity. Nebular emission also yields higher
\Qd\ measures.  Although the standard cluster tracks only reach \Qd=-0\fm3,
clusters with values up to -0\fm7 are observed in the LMC-SMC
\citep{pessev06}, which can explain the negative tail of \Qd.

\begin{figure*}
  \resizebox{0.98\hsize}{!}{\includegraphics{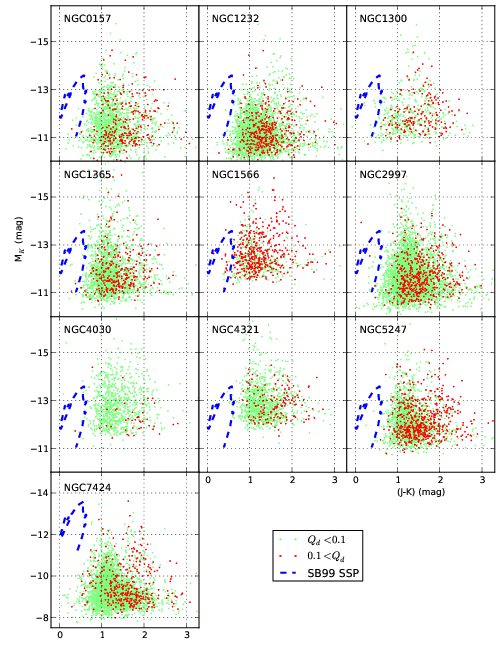}}
  \caption{Absolute \Ks-band magnitude \MK\ of all non-stellar sources with
    errors $<$0\fm3 as a function of their (J-K) index. The dashed line shows
    a SB99 SSP model with a mass of $10^5$\,\Ms.  The color code shows whether
    the \Qd\ index is above (red/dark) or below 0\fm1 (green/light). }
  \label{figa:jkall}
\end{figure*}

Color-magnitude diagrams for (J-K)--\MK\ are provided in Fig. \ref{figa:jkall}
with the standard SB99 SSP model as reference.  The younger population
(red/dark) is more prominent for high (J-K) values which are indicative of
higher extinctions.  As discussed in the main section, young clusters should
have lower intrinsic (J-K) values than older ones.  That they actually have
higher values suggests that they are highly obscured.  Double peaks are
clearly seen for NGC~157, NGC~1566, NGC~2997, and NGC~7424, while only
marginal ones are visible for NGC~1232 and NGC~5247.  A clear gap indicates a
rapid expulsion of dust from young clusters, while the lack of one could be
caused by both a slower expulsions and a wider spread in attenuation by dust.
The left edge of the distribution is fairly sharp and corresponds well to the
cluster track.  This indicates that many older clusters have low internal
extinction.

\section{Spatial distribution of complexes}
\begin{figure*}
  \resizebox{\hsize}{!}{\includegraphics{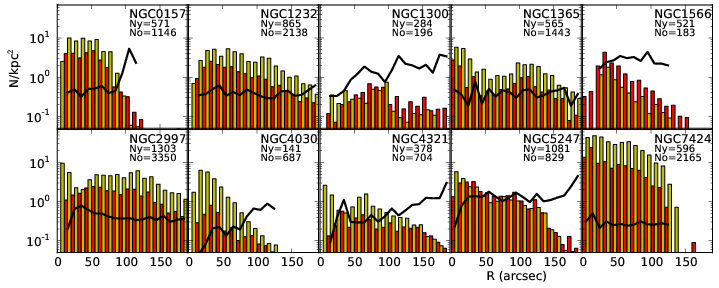}}
  \caption{Surface density of non-stellar sources with photometric errors
    $<$0\fm5. Total numbers of young objects, Ny, (with 0\fm1$<$Q) and older
    ones, No, (with Q$<$0\fm1) are listed for each galaxy.  Red (dark) bars
    represent young objects, while older ones are indicated in yellow (light).
    The ratio of young to old clusters is plotted as a full drawn curve. }
  \label{figa:nrhist}
\end{figure*}

The radial surface density of non-stellar sources is displayed in
Fig.\ref{figa:nrhist}, where red (dark) indicates younger and yellow (light)
older ones.  All galaxies show a steep decline in their outer parts, while
some (e.g. NGC~157, NGC~1365, NGC~2997, NGC~4321, NGC~5247, and NGC~7424) have
a relative flat part in their disks.  The distributions of younger and older
complexes follow each other well indicating that current and past SFRs display
little radial change.

\begin{figure*}
  \resizebox{0.97\hsize}{!}{\includegraphics{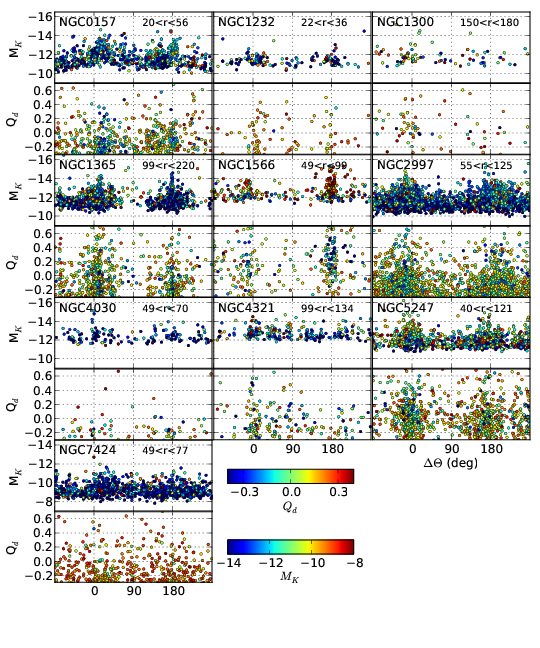}}
  \caption{Absolute magnitude \MK\ and \Qd\ index of non-stellar sources with
    errors $\sigma$(H-K)$<$0\fm3 as a function of their azimuthal distance
    $\Delta\theta$ from the phase of the m=2 FFT component.  The radial range
    used is indicated in the upper right corner and corresponds to the regions
    occupied by the main symmetric, two-armed spiral pattern.  Colors from
    blue to red indicate in the \MK\ plots the value of \Qd, while in the
    \Qd\ diagrams they show \MK\ as in Fig.~\ref{fig:mq2997}.  }
  \label{figa:mqall}
\end{figure*}

The locations of complexes relative to the spiral arms are given in
Fig.~\ref{figa:mqall}, which plots \MK\ and \Qd\ against the azimuthal distance
$\Delta\theta$ from the phase of the m=2 FFT component.  Only sources in the
radial range of the strong, grand-design spiral pattern (listed in the upper
right corner) and photometric errors $<$0\fm3 are shown.  The modulation of
the source density with azimuthal angle from the spiral arms is stronger for
the galaxies with strong, grand-design patterns such as NGC~157, NGC~1365,
NGC~1566, NGC~2997, and NGC~5247.  Both NGC~1232 and NGC~7424 have relatively
weak spiral perturbations, whereas the radial regions for NGC~4030 and NGC~4321
are located in their outer parts, which have less star-formation activity.  As
the Kolmogorov-Smirnov tests show, the shape of the distribution of
\Qd\ (i.e. age) does not change significantly from arm to inter-arm regions.
On the other hand, the excess of bright sources in the arms seems real.

\end{appendix}

\end{document}